# Decoupled algorithm for transient viscoelastic flow modeling and description of elastic flow instability

**Youngdon Kwon**[*a,b]

[a] *School of Chemical Engineering, Sungkyunkwan University,*
*Suwon, Gyeonggi-do 440-746, Korea*

[b] *Institute of Chemical Research, Kyoto University*
*Uji, Kyoto 611-0011, Japan*

---

[*] Corresponding author. Tel.: +82 31 290 7318; fax: +82 31 290 7330
*E-mail address:* kwon@skku.edu

**Abstract**

In the framework of finite element analysis we employ fast decoupled time integration scheme for viscoelastic fluid (the Leonov model) flow and then investigate strong nonlinear behavior in 2D creeping contraction flow. The algorithm of the $2^{nd}$ order is applicable in the whole range of the retardation parameter and shown to disclose convergence characteristics equivalent to the conventional method of corresponding order. In the analysis of steady solutions executed as a preliminary study, there exists upper convergence limit of available numerical solutions in this contraction flow, and it is free from the frustrating mesh dependence when we incorporate the tensor-logarithmic formulation [Fattal and Kupferman, J. Non-Newtonian Fluid Mech. 123, (2004) 281]. With adjustment of a nonlinear parameter in the Leonov model, 2 kinds of elastic fluid have been chosen for flow modeling such as highly shear thinning and Boger-type (weakly shear thinning and highly extension thickening) liquids. According to the type of such liquid property, the transient computational modeling has revealed qualitatively distinct flow dynamics of elastic instability. With pressure difference imposed slightly below the steady limit, current as well as conventional approximation scheme demonstrates fluctuating solution without approaching steady state for the shear thinning fluid. From the result, we may conclude that the existence of upper limit for convergent steady solution possibly implies transition to spatially as well as temporally varying flow field without steady asymptotic. When the pressure fairly higher than the limit is enforced at the inlet, the result expresses severe fluctuation of flowrate, oscillation of corner vortices and also asymmetric irregular stress wave propagation along the downstream channel wall. In addition, flow dynamics seems quite stochastic with almost no temporal correlation. For the Boger-type fluid, when the traction higher than steady limit is applied, the flowrate and corner vortices (and thus dynamic variables) exhibit periodic oscillation with flow asymmetry, while no fluctuation along the downstream wall has been observed. Both types of instability express purely elastic flow instability in this inertialess flow approximation described at least by the current set of equations. Verification of their relation with real unstable flow phenomena certainly requires further study involving mesh refinement, 3D flow simulation and realistic modeling of liquid employing e.g. relaxation time spectrum.

# 1. Introduction

Especially in highly nonlinear regime, elastic fluid like polymeric solution displays unique flow behavior, drastically different from Newtonian one and often against our common intuition. Rod climbing, tubeless siphon, extrudate swelling, large vortex formation [1,2,3,4], and purely elastic symmetry breaking [5] are the most representative of elastic flow phenomena observed in steady flow. In addition to these steady effects, there are characteristics known as "purely elastic instabilities" [6,7], the approximate description of which is main concern in this work. They are intrinsically nonlinear and possibly defined as complex (sometimes chaotic) flow phenomena accompanied by secondary flows or intense temporal as well as spatial fluctuations without steady asymptotic solely due to liquid elasticity. They occur even at low Reynolds number where inertial effect is too small to include in theoretical and numerical treatment.

Experiments, theoretical stability analyses and direct computational studies have been conducted to accumulate knowledge in this area of industrial as well as scientific significance. In order to achieve their realistic and efficient modeling, we here employ a fast and stable computational scheme for finite element modeling of transient viscoelastic fluid flow implementing a simple idea of decoupling. With this algorithm, we examine possible variation of instability mechanisms in contraction flow according to liquid property described by the Leonov constitutive equations. Thus we give brief overview of previous works mainly on numerical modeling rather than experimental discoveries.

During recent decades, important and essential progress has been made with finite element [8], finite volume [9], their hybrid method [10] and so on [11]. However in modeling highly elastic flow effects we still suffer from a variety of numerical breakdown often with uncertain origin. It is usually expressed as lack of convergence and loss of time evolution combined with violation of positive-definiteness of the configuration tensor. Preserving the positive-definiteness during computation is quite crucial, since its violation immediately incurs the Hadamard instability and eventually total breakdown of the computation procedure [12].

The degree of elasticity in flow is usually expressed in terms of the Deborah ($De$) or the Weissenberg number. It is defined as $De = \theta U / H_0$, where $\theta$ is the characteristic relaxation time of elastic liquid, $U$ the flow speed and $H_0$ the length scale. Hadamard or dissipative instability is an inherent characteristic of specific viscoelastic constitutive equations in the flow condition exceeding some critical limit of $De$ [13]. However even models proved globally stable (i.e. evolutionary in Hadamard and dissipative sense) exhibit numerical failure possibly due to unavoidable introduction of approximation error. In current author's opinion, rarity in direct computational description of elastically unstable phenomena reflects persistence of this difficulty. Recently definite remedy for this artifact has been introduced, and it is defined as matrix-logarithmic formulation of constitutive equations [14], here termed as tensor-logarithmic formulation. It strictly preserves the positive-definiteness just by setting in the new mathematical form, suggests breakthrough in flow modeling, but this may work only for the mathematically evolutionary models [15].

Time-dependent solutions of 2 or 3 dimensional problems are now quite a few with an advancement of computation power and techniques. The book and articles [8,9,10,11] provide a good review or recent results on various computational frameworks and thus here remark on previous works is not made. Instead we mention some recent results on the description of elastic instability in inhomogeneous flow field. Flow mark surface defect in injection molding process has been investigated by employing a linear stability analysis with basic information supplied by results of finite element analysis [16,17]. Even if this work provides information on criterion of instability and controlling its onset, the direct description of transient unstable flow has not been attained. Several years ago direct numerical description of unstable flow with an origin of elasticity has been reported [18,19], which expresses purely elastic symmetry breaking in cross and T-shaped channel flows. This year another considerably meaningful result has appeared, which numerically describes unstable flow dynamics in contraction flows [20]. These recent 2 classes of study describe flow asymmetry incurred by fluid elasticity, and besides the work on contraction flow expresses temporally fluctuating flow fields in the form of vortex oscillation and even back-shedding of vortices. They employed the finite volume method for spatial discretization applied essentially to the Oldroyd-B model.

In this study, we first introduce the set of total field equations for incompressible isothermal elastic flows. Before our discussing start-up flow problems, steady solutions free from mesh degeneration are obtained for creeping planar 4:1 contraction flow with sharp corner. The result suggests proper convergence of the scheme according to spatial refinement when the tensor-log transform is implemented in the formulation. Decoupled as well as conventional discrete time integration schemes are given in the Appendix. Even if the decoupled algorithm herein is suggested independently, its essential idea is the same with the one by D'Avino and Hulsen [21] except an additional intermediate step to deal with an extra nonlinear term in the Leonov model. Adjusting the nonlinear parameter in the constitutive equation, we exploit 2 qualitatively distinct liquid types such as highly shear thinning and Boger-type characteristics. After brief verification of the validity of the scheme from the viewpoint of accuracy and convergence at the same time, the time dependent contraction flow is then modeled. Analyzing the time-dependent flow behavior near steady convergence limit, we make an attempt to elucidate a puzzle implied by existence of such upper limit. Both the half (with symmetry boundary) and full domain problems are considered, and we demonstrate a solution in highly nonlinear regime which consists in the region far beyond the steady limit and expresses purely elastic instability. The results reveal quite distinct instability dynamics according to fluid quality such as degree of shear thinning or extension hardening.

## 2. Governing equations

In order to describe dynamic flow behavior of incompressible fluid, we first require the equations of motion and continuity

$$\rho\left(\frac{\partial \mathbf{v}}{\partial t} + \mathbf{v} \cdot \nabla \mathbf{v}\right) = -\nabla p + \nabla \cdot \boldsymbol{\tau}, \qquad \nabla \cdot \mathbf{v} = 0. \tag{1}$$

Here $\rho$ is the liquid density, $\mathbf{v}$ the velocity, $\boldsymbol{\tau}$ the extra-stress tensor and $p$ is the pressure. The gravitational force is neglected in the analysis and $\nabla$ is the gradient operator. When kinematic relation of the extra-stress is specified in terms of the constitutive model, the set of governing equations becomes closed for isothermal incompressible viscoelastic flows.

In expressing dynamic property of the elastic liquid, one version of the Leonov constitutive equations [4,22] is employed. Then it can be written into the following form:

$$\boldsymbol{\tau} = (1-s)G\left(\frac{I_1}{3}\right)^n \mathbf{c} + 2\eta s \mathbf{e}, \qquad \mathbf{e} = \frac{1}{2}\left(\nabla \mathbf{v} + \nabla \mathbf{v}^T\right), \qquad (n > 0, \ 0 \le s \le 1)$$

$$\frac{d\mathbf{c}}{dt} - \nabla \mathbf{v}^T \cdot \mathbf{c} - \mathbf{c} \cdot \nabla \mathbf{v} + \frac{1}{2\theta}\left(\mathbf{c}^2 + \frac{I_2 - I_1}{3}\mathbf{c} - \boldsymbol{\delta}\right) = \mathbf{0}. \tag{2}$$

Here $\mathbf{c}$ becomes the recoverable strain tensor that explains elastic strain in the Finger measure during flow. $\dfrac{d\mathbf{c}}{dt} = \dfrac{\partial \mathbf{c}}{\partial t} + \mathbf{v} \cdot \nabla \mathbf{c}$ is the total time derivative of $\mathbf{c}$,

$\dfrac{d\mathbf{c}}{dt} - \nabla \mathbf{v}^T \cdot \mathbf{c} - \mathbf{c} \cdot \nabla \mathbf{v}$ the upper convected time derivative, $G$ the modulus, $\theta$ the relaxation time, $\eta = G\theta$ the total viscosity that corresponds to the zero-shear viscosity and $s$ is the retardation parameter (ratio of retardation to relaxation time) that may qualitatively specify the contribution of viscous solvent. The tensor $\mathbf{c}$ reduces to the unit tensor $\boldsymbol{\delta}$ in the stationary state and this also serves as the initial condition in the start-up flow from the rest.

$I_1 = \text{tr}\,\mathbf{c}$ and $I_2 = \text{tr}\,\mathbf{c}^{-1}$ are the basic first and second invariants of $\mathbf{c}$, respectively, and they coincide in planar flows. Due to the characteristic valid for the Leonov model, the third invariant $I_3$ has to fulfill the incompressibility condition, $I_3 \equiv \det \mathbf{c} = 1$, which reflects feasible physical meaning [23]. In addition to the linear viscoelastic parameters $G$, $\theta$ and $s$, there is a nonlinear constant $n$ ($n>0$), which can be determined from simple shear and uniaxial extensional flow experiments certainly in the nonlinear regime. It controls the strength of shear thinning and also extension hardening of the non-Newtonian liquid. The total stress tensor is obtained from the elastic potential $W$ based on the Murnaghan's relation. Since the extra-stress is invariant under the addition of arbitrary isotropic terms, when one presents stress results it may be preferable to use instead, e.g. $\tau = (1-s)G\left(\dfrac{I_1}{3}\right)^n (\mathbf{c} - \delta) + 2\eta s \mathbf{e}$ in order for the stress to vanish in the rest state.

The essential idea presented by Fattal and Kupferman [14] in reformulating the constitutive equations is the tensor-logarithmic transformation of $\mathbf{c}$ as follows:

$$\mathbf{h} = \log \mathbf{c} . \tag{3}$$

Here the logarithm operates as the isotropic tensor function, which implies the identical set of principal axes for both $\mathbf{c}$ and $\mathbf{h}$. In the case of the Leonov model, $\mathbf{h}$ becomes another measure of elastic strain, that is, twice the Hencky elastic strain. While $\mathbf{c}$ becomes $\delta$, $\mathbf{h}$ reduces to $\mathbf{0}$ in the rest state. In terms of $\mathbf{h}$, the incompressibility relation ($\det \mathbf{c} = 1$) is equivalent to

$$\operatorname{tr} \mathbf{h} = 0 . \tag{4}$$

In the following we introduce the following dimensionless set of variables:

$$\frac{x_1}{H_0} = x, \ \frac{x_2}{H_0} = y, \ \frac{t}{\theta} \to t, \ \frac{\theta}{H_0} v_i \to v_i, \ \frac{\tau_{ij}}{G} \to \tau_{ij}, \ \frac{p}{G} \to p,$$

$$Re = \frac{\rho U H_0}{G \theta}, \ De = \frac{\theta U}{H_0} , \tag{5}$$

where $H_0$ is the characteristic length scale of the flow channel depicted in Fig. 1, and $U$ is some representative flow speed. $Re$ and $De$ are Reynolds and Deborah numbers, respectively. We also define the dimensionless average flowrate $\bar{v}$ as

$$\bar{v} = \int_0^1 v_1 dy \quad \left( = \frac{\theta}{H_0{}^2} \int_0^{H_0} v_1 dx_2 \quad \text{in dimensional form} \right) . \tag{6}$$

When we assign $U$ as the downstream mean axial speed, $\bar{v}$ coincides with $De$. On the other hand, in the problem of full flow domain depicted in Fig. 1(b) where there possibly occurs symmetry breaking, the average is taken over the whole range of $y$, i.e., from -1 to 1.

In the 2D flow problems formulated in the tensor-log form, the Leonov model possesses its own advantage due to the simplicity represented as Eq. (4). The final set of the Leonov constitutive equations in the tensor-log or **h**-form has been obtained in Ref. [24] as follows:

$$\frac{\partial h_{11}}{\partial t} + v_1 \frac{\partial h_{11}}{\partial x} + v_2 \frac{\partial h_{11}}{\partial y} - \frac{2}{h^2}\left(h_{11}^{\,2} + h_{12}^{\,2} h \frac{e^h + e^{-h}}{e^h - e^{-h}}\right)\frac{\partial v_1}{\partial x}$$

$$- h_{12}\left[\frac{h_{11}}{h^2}\left(1 - h\frac{e^h + e^{-h}}{e^h - e^{-h}}\right) + 1\right]\frac{\partial v_1}{\partial y} - h_{12}\left[\frac{h_{11}}{h^2}\left(1 - h\frac{e^h + e^{-h}}{e^h - e^{-h}}\right) - 1\right]\frac{\partial v_2}{\partial x}$$

$$+ \frac{e^h - e^{-h}}{2h}h_{11} = 0,$$

$$\frac{\partial h_{12}}{\partial t} + v_1 \frac{\partial h_{12}}{\partial x} + v_2 \frac{\partial h_{12}}{\partial y} - \frac{2h_{11}h_{12}}{h^2}\left(1 - h\frac{e^h + e^{-h}}{e^h - e^{-h}}\right)\frac{\partial v_1}{\partial x}$$

$$- \left[\frac{1}{h^2}\left(h_{12}^{\,2} + h_{11}^{\,2} h \frac{e^h + e^{-h}}{e^h - e^{-h}}\right) - h_{11}\right]\frac{\partial v_1}{\partial y} - \left[\frac{1}{h^2}\left(h_{12}^{\,2} + h_{11}^{\,2} h \frac{e^h + e^{-h}}{e^h - e^{-h}}\right) + h_{11}\right]\frac{\partial v_2}{\partial x}$$

$$+ \frac{e^h - e^{-h}}{2h}h_{12} = 0. \tag{7}$$

Here $h = \sqrt{h_{11}^{\,2} + h_{12}^{\,2}}$ is the only 1 independent eigenvalue of **h** in 2D flow (the other eigenvalue is $-h$). The relation between $c_{ij}$ and $h_{ij}$ is expressed by

$$c_{11} = \frac{1}{2}\left[e^h + e^{-h} + \frac{h_{11}}{h}\left(e^h - e^{-h}\right)\right], \quad c_{22} = \frac{1}{2}\left[e^h + e^{-h} - \frac{h_{11}}{h}\left(e^h - e^{-h}\right)\right] \text{ and}$$

$$c_{12} = \frac{h_{12}}{2h}\left(e^h - e^{-h}\right). \tag{8}$$

Together with Eqs. (1), Eqs. (7) constitute a complete set to describe isothermal incompressible planar viscoelastic flow. However due to their mathematical form presented in Eqs. (7), artificial numerical difficulty may arise. Including the rest state, during flow the incident of $h$=0 (it means **h = 0**) may occur locally, e.g. along flow symmetry line. Then the coefficients of $\partial v_i/\partial x_j$ and $h_{ij}$ in Eqs. (7) become numerically indeterminate. However appropriate asymptotic relation for vanishing $h$ can be easily obtained and given in Ref. [24].

In this 2D analysis, the incompressibility condition (4) reduces to $h_{11} = -h_{22}$, $h_{33} = 0$ and thus the viscoelastic constitutive equations add only 2 extra unknowns such as $h_{11}$ and $h_{12}$ to the total set of variables.

## 3. Numerical method

The temporal integration methods are given in Appendix A. IE1 and CN2 are the implicit Euler and Crank-Nicolson schemes, while DC1 and DC2 are the decoupled ones of order 1 and 2, respectively. Except for specific occasions computed with IE1 and CN2, all the computations are performed with DC2 in this work.

In planar 2D flow, standard finite element analysis is adopted as a basic framework. Stabilizing techniques such as streamline-upwind/Petrov-Galerkin (SUPG) and discrete elastic viscous stress splitting (DEVSS) methods are also implemented. Nowadays they are quite routinely applied to finite element viscoelastic flow computation, and thus their detailed meaning is not explained and may be found in Refs. [8,11,25]. Quadratic polynomial interpolation is employed for velocity and elastic strain $h_{ij}$, while linear interpolation is used for pressure and DEVSS variables.

Problem domain with specific boundary conditions is illustrated in Fig. 1. In planar 4:1 contraction flow geometry, we designate no slip on the wall (and symmetry on the centerline for a half-domain problem (a)), but 2 more additional boundary line segments exist at the inlet and outlet of the channel. In this study, traction forces are appointed to induce the flow, the difference of which corresponds to the applied pressure at the entrance. We suppose that in comparison with fully developed inflow condition this traction boundary may readily allow occurrence of elastic flow instability due to flowrate fluctuation permitted under fixed pressure difference. First, all the components of traction vanish at the outlet. On the other hand, in the flow direction ($y$-axis) at the inlet, the constant finite traction force is applied in terms of dimensionless value $t_y/G$ where $t_y$ the surface traction (force per unit area) in $y$-direction. In the $x$ direction, we set the boundary free from transverse traction.

We also need to specify values of **h** at the inlet boundary, which play the role of initial conditions at the beginning of the characteristic curves, but they are unknown and have to be determined in principle from the previous flow history.

Here we simply assign zero for all components of **h**-tensor (i.e. stress free condition) at the inlet, which may necessitate some discussion to justify this rather arbitrary specification. This type of boundary means that the region outside the inlet boundary is considered as a zero stress reservoir just like pressure or heat reservoir in thermodynamics. In practice, the length of the upstream channel has to be long enough if one wants to eliminate the effect of this stress-free boundary and this is why we employ somewhat long upstream channel in contrast to the downstream (Fig. 1). From the numerical experiment we could observe that fully developed flow is immediately attained near the inlet whenever the steady flow state is reached. On the contrary the fully developed state in temporally varying flow situation cannot be realized near the inlet and even becomes meaningless under the problem definition herein.

In the start-up flow problem, initial values for velocity and **h**-tensor are also necessary. Regarding this, assigning traction boundary rather than developed flow condition bestows its own advantage. We can instantly specify arbitrarily large inflow traction to incur high *De* flow, and furthermore onset of flow from the rest state in the whole domain is also permissible without any logical inconsistency. Therefore we simply designate the values corresponding to the rest state (all 0 values) as initial data.

In the solution of 2D contraction flow, 8 types of meshes with unstructured Delaunay triangular elements are constructed. Table 1 presents mesh details such as side length of the smallest element and the total numbers of elements and unknowns for half and full domain problems. The mesh of the full domain problem is constituted with attaching to the half domain its exact mirror image and thus becomes perfectly symmetric along the centerline. Among them 2 different sets of 5 meshes are chosen for the problems with $n$=0.1 and $n$=2, where the nonlinear parameter $n$ is characterized in the following section. Especially in the case of $n$=2, the set of rather coarse meshes is constituted since the transient flow modeling requires very fine time stepping and thus the computation becomes quite demanding.

In viscoelastic flow computation for domain involving geometric singularities, there is one persisting dilemma represented as numerical degeneration according to spatial mesh refinement.    We first investigate this problem on steady flow computation and then make an attempt to elucidate any possible implication supported by existence of computational convergence limit.    Fig. 2 illustrates a partial view of Mesh 6 around contraction together with steady solution of shear stress $\tau_{12}$ under dimensionless input traction of 50.    As a matter of course, the smallest elements are attached to the contraction corner.    As for the analysis of time dependent flow, rather coarse meshes such as Mesh 1 and 3 have been mainly applied since even with the decoupled method computation load required for refined meshes is still beyond the capacity we can manage.    Note that we primarily examine the mesh dependence of steady solutions, with an emphasis on the highest $De$ or traction with proper convergent solutions.    Mesh convergence test of the transient solutions will be very limited in this study.    Thus especially in high $De$ flow we examine the possibility of numerical description of elastic instability, presenting data perhaps in lack of quantitative accuracy.

For numerical calculation, we need to specify values of linear and nonlinear parameters such as $s$ and $n$, for which admissible ranges are $0 \leq s \leq 1$ and $n > 0$.    In order to suppress solvent viscosity effect and thus augment elastic character in the model fluid, we assign small value of the retardation parameter $s$.    However setting $s$=0 in inertialess flow induces infinity of mainstream velocity at the moment of application of pressure difference, and thus we have chosen $s$=0.001. For the nonlinear parameter $n$, there exists some mathematical stability requirement [13].    For so-called Hadamard stability corresponding to strong ellipticity in elasticity, $n$ has to be nonnegative.    However in the case of neo-Hookean potential, the model becomes dissipative unstable.    In other words, it may show blow-up instability for $n$=0 when mixed stress-strain history is assigned.    The problem in this study is exactly the case that may incur the dissipative instability, since we specify the traction (total stress) boundary instead of velocity profile at the inlet.    Therefore if the applied inlet pressure exceeds some critical value with the neo-Hookean potential employed, the scheme will exhibit loss of evolution and become divergent.    This appoints slightly stronger restriction of $n$>0 and thus eliminates a chance of using the neo-Hookean potential.    Thus herein we set $n$=0.1 and 2, with which the constitutive equation is defined to be globally stable [13]. The reason of choosing these two values is further explained in the next section.

# 4. Results

Before presenting results in 2D flow, we discuss the behavior of the Leonov model in simple flows. The constitutive equation employed in the current study is identical to the one in the work [26], when $b(I_1, I_2) = 1$ and the elastic potential $W = \dfrac{3G}{2(n+1)}\left[\left(\dfrac{I_1}{3}\right)^{n+1} - 1\right]$. Thus the simple flow characteristics of the model except the characterization of $n$ are well illustrated in the referred paper. Therefore the dependence of steady flow response only on the change of $n$ is here explained.

The simple flow curves are shown in Fig. 3, where strain rates and viscosities are scaled with the relaxation time and zero-shear viscosity ($\eta = G\theta$), respectively. Steady shear response exhibits qualitative change at $n=2$ where it shows weak shear thinning and the dimensionless viscosity at high shear approaches the asymptotic value of $(4/9)(1-s) + s$ as depicted in Fig. 3(a). When $n<2$, the model describes behavior of shear thinning with asymptotic Newtonian viscosity $s$, whereas shear thickening emerges for $n>2$. Hence we choose $n=0.1$ as the representative of shear thinning liquid and $n=2$ for which the model roughly resembles the property of the Boger fluid.

Fig. 3(b) portrays extension hardening behavior in every case of $n$, even though the higher value significantly intensifies it. Since the extensional flow that may occur in contraction flow of the current work is not uniaxial but planar, the steady planar extensional viscosity as a function of extension rate is also described in Fig. 3(c). Even though there is some quantitative distinction, the overall characteristic is the same with the uniaxial one.

## 4.1. Steady contraction flow

We first of all introduce the result of steady contraction flow modeling. For several values of $n$, Fig. 4 shows dependence of the dimensionless flowrate in the measure of the Deborah number upon the pressure difference (dimensionless traction) computed for Mesh 3. Except for the shear thickening case of $n=3$, every curve for shear thinning fluid ($0<n\leq2$) displays higher flowrate than the Newtonian straight line, and as anticipated, higher thinning results in higher flowrate. Thus at least in determining flowrate for planar contraction ratio 4:1, it seems that the shear property dominates over the extensional one.

When applied flow condition in terms of the Deborah number or traction force exceeds some critical limit, many viscoelastic constitutive equations are proved to yield some ill-posed behavior, which eventually results in the loss of evolution [12,13]. From the mathematical stability analysis, it is proved that there exist only a few globally stable rheological equations, one of which is the Leonov model [13]. In addition to this mathematical stability, one may require another property of constitutive models in order to elucidate intimation possibly implied by the limitation of convergence in numerical steady solutions. Due to the incompressibility relation (4), the Leonov model becomes the simplest of all constitutive equations in **h**-form, and this simplicity makes the closed formulation (7) possible. Just dropping the time differential terms offers the set of nonlinear equations, solution of which directly provides steady values of rheological variables without computationally passing through transient state. Within author's knowledge, steady 2D numerical solutions obtained in this straightforward manner with tensor-log formulation are very limited (recently steady results have also been presented by generalized strong coupling [27] and with a monolithic approach [28]). Directly obtaining steady solutions not in **c** but in **h** formulation is quite essential in this study to elucidate cause of limit in steady computation. Henceforth we present steady results for the 4:1 contraction flow problem.

In modeling viscoelastic flow especially with a singular corner in a domain, we have been faced with one concomitant obstacle, that is, breakdown of numerical scheme more severely with mesh refinement [29]. In other words, as one increases the degree of spatial refinement usually to improve accuracy, the limit Deborah number over which convergent steady solution cannot be obtained decreases quite rapidly. This kind of frustrating contradiction has been existent even for the globally stable models when **c**-form or stress-type constitutive equations are implemented in computational approximation. The loss of numerical evolution is usually accompanied by violation of the positive-definiteness of the **c** tensor as its precursor. Even though preserving the positive-definiteness has been proved in some well-defined analytical situation [30,31,32], indispensable numerical error frequently results in its violation and eventually whole degradation of the procedure at high *De*.

Main advantage of the tensor-log formulation in discrete approximation consists in the strict preservation of the positive-definiteness, which can be easily understood from the relation $\mathbf{c} = \exp(\mathbf{h})$, the inverse of Eq. (3). Thus no matter what value **h** takes in the real space, **c** is always positive-definite. In the principal axes of **c** the range 0~1 of eigenvalues is equivalent to 1~∞ for the Leonov model (in 2D the eigenvalues are *c*, $1/c$, 1), and by the tensor-log transformation those ranges are recast to −∞~0 and 0~∞ in **h**-form. Therefore polynomial interpolation of **h** rather than **c** tensor endows more consistent approximation in the whole positive-definite range of **c**. Also steep viscoelastic boundary layer present in high Deborah number flow seems better resolved with this formulation [27].

In this work, 2 different sets of meshes for *n*=0.1 and 2 are employed with increasing degree of spatial refinement illustrated in Table 1, where the limit dimensionless tractions (pressure difference) and the corresponding steady Deborah numbers (flowrate) are also listed. Here, in reality, the solution with *n*=0.1 is obtained for the half domain problem whereas the full domain problem is solved for *n*=2. However computing both for several meshes, we have confirmed that half and full geometry problems yield identical results due to no symmetry breaking under steady flow assumption in the full domain. Even though it certainly displays some scatter of limit values especially for *n*=2, one can hardly notice distinct sign of disastrous degradation of the computation scheme according to mesh refinement. One example of shear stress solution directly obtained from the steady equations slightly below the critical limit ($\Delta p = t_y / G = 50$, $n = 0.1$, Mesh 6) is illustrated in Fig. 2 as a contour plot. Note that with larger value of the retardation parameter, e.g. *s*≥0.5, the convergent steady solution may be obtained for $De$ as high as thousands. However in this case highly elastic phenomena like purely elastic instability or asymmetry may be suppressed by the excessive viscous dissipation at the same $De$.

The dependence of steady flowrate upon traction is demonstrated in Fig. 5. One should expect linear dependence of the flowrate upon $\Delta p$ for Newtonian fluid with no inertia, which is illustrated in Fig 4 as a solid line. Here stiff upturn of the curve is depicted for $n$=0.1 (Fig. 5a), which results from the characteristic shear thinning behavior expressed by the Leonov model. In the case of $n$=2, after slight initial convexity caused by weak shear thinning in the shear rate range between 1 and 10 (Fig. 3a), the dependence becomes almost linear due to asymptotically constant shear viscosity at high shear.

Mesh dependence of flowrate can be observed in the insets of Fig. 5, where finer mesh describes higher flowrate or $De$ and its mesh dependence becomes stronger at high traction. Thus in high $De$ flow the degree of spatial approximation rather seriously affects the accuracy of numerical solution. This sort of sensitivity on discretization in nonlinear dynamics is quite common and becomes extremely hard to overcome when quantitatively accurate results are desired. However in comparison with the curves for $n$=0.1, the solution with $n$=2 exhibits weaker mesh dependence.

As an intermediate conclusion, in terms of the tensor-log formulation for discrete approximation of highly elastic steady flow one can overcome the persistent obstacle expressed by the mesh dependence of the upper convergence limit. Nevertheless there still remains limitation for attaining numerical steady solution in contraction flow problems such as $De_c$=53~55 for $n$=0.1 and $De_c$=5~7 for $n$=2. In the following we analyze these steady solutions in another aspect, solving a time dependent problem near but still below the upper limit, and then determine if they really represent steady approximation described by the constitutive model.

### 4.2. Transient contraction flow of highly shear thinning fluid

In this subsection we examine time-dependent flow of shear thinning liquid

($n$=0.1).   Solutions of the weak form for Eqs. (1) and (7) implemented with IE1, CN2 and DC2 schemes on Mesh 3 are illustrated in Fig. 6 for half domain under the pressure difference $\Delta p$=30.   Here the applied pressure is well below the upper steady limit for all meshes.   The dimensionless time step $\Delta t$ is fixed as $10^{-4}$ or $5\times10^{-5}$.   As illustrated by all the curves, the flowrate slowly reaches its steady value following some transient response.   When we regard the CN2 solution of $\Delta t$=$5\times10^{-5}$ as the most accurate one, the 2nd order schemes such as CN2 and DC2 yield results with higher accuracy as one should expect.   In the inset of Fig. 6, actually the curve of DC2 ($\Delta t$=$10^{-4}$) overlaps with the most accurate one of CN2 ($\Delta t$=$5\times10^{-5}$).   On the other hand, the lower order method IE1 exhibits slightly earlier transient than CN2 or DC2 does.   This presents one evidence required for the accuracy of DC2 equivalent to conventional method CN2 in the medium range of nonlinearity.

Fig. 7 portrays overall time variation of flowrate described by CN2 and DC2 schemes with Mesh 5 for $\Delta p$=50 which still resides in the region of available steady solution for all 5 half domain meshes.   The time step is varied[1] in the range 1~$2.5\times10^{-4}$.   Quite unexpected outcome can be observed from the curves computed in this study.   Even after initial transient response, the flowrate does not approach the steady state value, and instead it exhibits fluctuation with some mixed spectral mode.   Hence the solution obtained under the steady flow assumption is shown to be a fictitious one.

Even if both the algorithms exhibit similar overall behavior, at $t$>1.5 quantitative discrepancy between two results can be observed in the figure inset. We interpret this not as the accuracy problem of DC2 but as typical behavior of approximate solution existent in nonlinear mechanics known as "sensitivity to initial conditions".   In other words, at every step of discrete time approximation we are destined to introduce tiny error, the magnitude and sign of which depend on specific numerical algorithm.   Such small variation in the nonlinear dynamical system consequently induces quantitatively different solutions, and the magnitude of their inconsistency increases with time. Thus such discrepancy and eventual lack of correlation between two solutions are inevitable in this kind of high nonlinearity.

---

[1] Even though the time step varies, the same sequence of different step sizes is applied for both algorithms.

Under the same condition ($n$=0.1 and $\Delta p$=50), the flowrate is computed for 4 different meshes in Fig. 8. First, the finer the mesh becomes, the higher value the flowrate achieves, and similar trend has already been observed in the steady flow curves in Fig. 5. Most remarkably, while for the coarse domains such as Mesh 1 and 3 the flowrate attains its steady state, for Mesh 5 and 6 it never becomes steady and exhibits fluctuation after some initial transients. For example, for Mesh 3 the flowrate ($De$) achieves its steady value 49.7 when $t$>4.7. However in the case of Mesh 6 even though the flowrate computed under steady approximation is 50.7, it always varies with time and its average value in the range $t$=3~12 increases to 53.64. In our opinion, this kind of flow transition occurs since the introduction of elastic flow instability enables more effective mechanism of relaxing high elastic energy accumulated during the flow. In addition, inevitable numerical diffusion existent either in start-up flow modeling with coarse mesh or in steady computation makes the asymptotic steady solution available. At this point one may conclude the following: the convergence limit in steady flow modeling is associated with the elastic transition from asymptotically steady to temporally fluctuating flow field, but this limit value does not exactly coincide with the transition point, which consists in somewhat lower region. The critical value of traction or the Deborah number over which the elastic flow instability occurs, is dependent upon the degree of spatial discretization and thus for its precise estimation very fine structure has to be employed. Here we do not pursue further to determine this transition point, since its determination is too demanding in computational load.

From now on we observe this unstable flow phenomena in more detail. Fig. 9 illustrates streamlines and isotropic pressure contours for the meshes and flow conditions same as in Fig. 8 at the time instant of $t$=5. First, in Fig. 9(a) one can examine the solution in the flow domain near the contraction corner. Among the solutions, slight difference can be observed like higher pressure gradient at the vortex detachment point (at the wall and $y \approx -6.5$) and smoother pressure variation near the reentry corner for finer mesh. However overall behavior does not show quite noticeable distinction. Fig. 9(b) where solution in the downstream channel ($5 < y < 15$) is portrayed, expresses some qualitative change with mesh refinement especially in pressure. While solutions of Mesh 1 and 3 exhibit only the smooth variation along the flow direction (since the steady state has been reached at $t=5$ for both solutions), for Mesh 5 and 6 the results disclose spatial pressure fluctuation along the wall. This means that description of such type of elastic instability at the current flow condition calls for spatial refinement enough to resolve the fine structure of fluctuation.

At 2 separate time instants ($t=4.275$ and $4.575$), the flow field is again shown in Fig. 10 for Mesh 6. 2 illustrations in Fig. 10(b) and (c) correspond to solutions at instants marked with small circles on the flowrate curve of Fig. 10(a). Thus left and right figures in both (b) and (c) represent computed flow fields near the minimal and maximal flowrates, respectively. Even though the difference is not so drastic, fluctuation in terms of smaller at $t=4.275$ and larger vortex at $t=4.575$ is expressed in Fig. 10(b). Even if the direction of propagation cannot be recognized, at least temporal variation of pressure oscillation along the wall can also be examined in Fig. 10(c).

Hereafter we consider the flow under much severer condition of $\Delta p=90$. Thus the applied traction is much higher than the steady convergence limit for all meshes. The solutions of overall flowrate, streamlines near contraction and streamlines with contour of elastic potential along the narrow channel are represented in Fig. 11, 12 and 13, respectively. The CFL condition states as follows: as the mesh gets finer or flow speed becomes higher, smaller time step has to be implemented in order to keep the Courant number small [33]. Owing to this stringent requirement, here only the coarse one, Mesh 3 is employed to relieve heavy computational load.

Fig. 11 illustrates results generated with the 2nd order integration methods. While for DC2 the temporal step size was fixed as 0.0002, for CN2 only up to $t=0.5$ the same step was applied and then gradually decreasing ones were employed to achieve convergence as far as possible. Nevertheless in the case of CN2, the stable temporal progress has been made only until $t$=0.7685266 even with $\Delta t$=10$^{7}$. On the other hand with DC2 the computation continued until $t$=5 and then stopped due to no further interest. At this point, the origin of loss of convergence with CN2 is not evident at all to the current author, and the successful numerical computation seemingly needs more careful control of time stepping.

In this highly nonlinear case, the flowrate exhibits vigorous fluctuation, and in the inset when $t$>0.54 we can observe noticeable deviation of DC2 solution from that of CN2 again due to the initial data sensitivity or caused by nonevolutionary character of CN2. The average flowrate in the range of $t \in (1, 5)$ reaches as high as 341.4 and the standard deviation is computed to be 23.77 in the Deborah number measure. In order to associate this result with real polymer flow, caution has to be given against its impetuous interpretation. There is e.g. one type of flow instability called "melt fracture", the origin of which is hardly clear yet. If it results from stick-slip on the channel wall, then the result shown here has nothing to do with it since in this case completely different boundary condition has to be assigned at the wall. On the other hand, the experimental observation performed in the microfluidic device [34,35] may share the origin of instability same with the current results, although the flow geometry is different and full 3D computation is required for realistic description of elastic turbulence and turbulent mixing.

In Fig. 12 streamlines at 4 time instants near contraction are drawn for $\Delta p$=90, where one can examine oscillation of the corner vortex. While its center displays distinct movement, its overall size varies little. At the same time instants and along the downstream channel, streamlines and contour of elastic potential are depicted in Fig. 13. The elastic potential of the Leonov model can be understood as the elastic energy accumulated during the flow. Its spatially fluctuating concentration near the wall is demonstrated and it roughly defines the viscoelastic boundary layer much thicker than that in Fig. 10(c). One can as well observe temporal fluctuation of the potential as variation along the wall in the boundary layer. Careful examination also shows time-dependent wavy streamlines near the wall, whereas in the main stream apart from the wall they are rather straight with negligible alteration. If the solution here is ever associated with real extrudate irregularity, this wavy streamline as well as the overall flowrate fluctuation is expected to cause gross distortion like melt fracture.

In Fig. 14 the temporal variation of overall flowrate is presented together with its FFT spectrum, calculated for data only in $3<t<5$, where we do not account for the data in $0<t<3$ to keep initial transients from consideration. The flowrate shows quite complicated fluctuation in these rather highly nonlinear situations. Even though direct comparison between FFT spectra as well as flowrates may not be justified due to different meshes employed, the higher boundary traction in the case of Fig. 14(b) enhances the complexity of temporal variation of flowrate in spite of the lower spatial resolution. This can also be examined directly from the FFT spectra, where higher harmonics build up as the traction increases.

Until now we have examined only the solution for the half domain postulated by the symmetry of flow dynamics along the channel centerline. However in the highly nonlinear situation with severe temporal and spatial fluctuation of flow field that we have just observed, such synthetic symmetry condition may no longer be valid. Thus in the following, the problem in the full 2D domain is taken into account but with arduous sacrifice of computational economy.

In comparison with the problem of half flow domain, that of full domain demands much higher computational resources due to the increased number of variables and especially larger bandwidth of the matrix to be inverted. This again limits us to the analysis only for rather coarse spatial resolution.

Fig. 15(a) exhibits the flowrate as a function of time for Mesh 1 and 3 under $\Delta p$=80, which again exceeds the limit value of convergent steady solutions for all meshes employed. After some initial transients it heavily fluctuates for both cases without reaching steady state. More refined mesh results in higher flowrate as the case of steady or transient (Fig. 8) solutions and also allows finer structure of time variation as revealed in the figure inset. FFT spectra for the data in $t$=1.5~3.5 are computed in Fig. 15(b). As is the case of half domain problem with $\Delta p$=50 and 90 in Fig. 14, it is hard to determine some definite mode of periodicity or characteristic harmonics and the behavior looks quite stochastic.

For Mesh 3, higher power of the spectrum explains larger amplitude of the fluctuation and distinct modes exist at higher frequency.

Now we analyze the distribution of dynamic variables at some fixed time instants in order to look into phenomena in the full domain. In the region near contraction entry, streamlines at 2 time instants $t$=3.024 and 3.062 are depicted in Fig. 16(a), while isotropic pressure contour with streamlines is drawn for the downstream channel flow in Fig. 16(b) additionally at $t$=3.094. If one recognizes change in location of the center of corner vortices, their fluctuation is distinctly manifested, even though the overall size does not show noticeable alteration. However breaking of symmetry is hardly present in Fig. 16(a), whereas there appears intense symmetry breaking in the narrow downstream channel to be discussed hereafter.

All the results at 3 different time instants in Fig. 16(b) display conspicuous temporal as well as spatial fluctuation of pressure along the wall, of which the similar phenomena have already been found in the half domain solutions of Figs. 10(c) and 13. Such spatial pressure gradients again propagate along the wall in the flow direction. In addition to this heavy instability, clear asymmetry also intensifies the complexity of flow dynamics. In addition, careful examination of streamlines similarly reveals wavy locus near the wall. The overall flowrate variation combined with asymmetry and fluctuation along the wall will assuredly incur severe distortion of output.

### 4.3. Transient contraction of Boger-type fluid

Now, we alter the value of the nonlinear parameter to $n$=2, with which the viscoelastic liquid possesses properties almost like the Boger fluid. For this problem we apply temporally varying inlet boundary traction depicted in Fig. 17, that is, it takes constant values from 300 to 400 with step increase, all of which exceed the convergence limit in steady solution. The reason for employing such sequence of step tractions is as follows. For this highly extension thickening (at the same time, weakly shear thinning) fluid, very careful time stepping turns out to be important for stable computation especially in the initial transient state.

In turn, when we apply higher boundary traction, utilizing the solution corresponding to slightly lower traction as initial condition effectively saves the computation time, since it alleviates the intensity of initial step variation and shortens initial transient period.　Then we may be able to identify any change of flow dynamics according to the variation of input traction without passing through whole process of start-up flow from the rest for each condition.

　　To begin with, effect of temporal step size is estimated.　In Fig. 18, the red and black curves denote flowrates computed with $\Delta t$=1.25×10$^{-5}$ and 0.5×10$^{-5}$, respectively (Mesh 1), where the applied traction is 300.　In Fig. 18(a) severe oscillation of flowrate appears when a larger step size is employed (red curve), however it disappears when sufficiently small step is used (black curve). D'Avino and Hulsen [21] have also observed and reported the same type of numerical artifact.　One peculiar feature of this decoupled method is that such erroneous oscillation does not always result in eventual failure of computational convergence and after some while it may vanish as shown on the red curve near $t$=1.147 of Fig. 18(a).　Such artificial fluctuation seems to introduce serious numerical diffusion, which in turn erroneously dissipates elastic energy accumulated during the flow.　However after some duration enough for restoration of dissipated elastic energy, the main feature of the flow dynamics recovers.　Actually Fig. 18(b) demonstrates such recovery since both curves express quantitatively identical behavior only with some phase difference (this Boger-type fluid exhibits perfectly periodic flow behavior, as we will examine in more detail from now on).　On the other hand, the flow problem for the highly shear thinning fluid ($n$=0.1) has not shown this type of artificial oscillation at least in our computation regime.　Thus the control of time step seems more crucial for the problem of the Boger-type fluid flow.

　　For the same flow condition, we have solved the half (red curve) as well as the full (black curve) domain problem with fixed $\Delta t$=0.5×10$^{-5}$ and then the temporal variation of flowrates is compared in Fig. 19.　Both curves coincide well until $t$=1.27, however after that instant discrepancy becomes noticeable.　Even though perfectly periodic state has not been reached yet, one can see in Fig. 19(b) that the interval between successive peaks for the half domain is almost double of that for the full domain.   When we inspect the flow behavior, for a short time period of $1<t<1.2$ the full domain solution exhibits symmetric corner vortex fluctuation.   Then when $t>1.25$, symmetry breaking starts to appear in the form of asynchronous undulations displayed by right and left vortices.   When the time of flow passes sufficiently, the flow dynamics becomes perfectly periodic.   In this case, one peak of the black curve in Fig. 19(b) corresponds to a half period with expanding right and contracting left corner vortices or vice versa.   Then in the following peak the rest half period is completed in the exact mirror image of vortex dynamics.   Thus in the full domain problem 1 period consists of 2 peaks in flowrate, while 1 peak represents the whole cycle for the half domain.   The split of peaks (and reduction in amplitude as a result) shown in the black curve from $t=1.31$ corresponds to the appearance of flow asymmetry in the form of temporal separation of vortex motions.

In reality, the spatial resolution employed for this problem may seem too coarse to draw decisive conclusion regarding viscoelastic flow dynamics. However from the result of steady flow analyses, we have seen that in this case of Boger-type fluid the solution is rather insensitive to the refinement of mesh. Thus we expect the analysis performed in this study still endows useful and correct information.   Furthermore, with finer Mesh 3 we have also observed onset of similar vortex fluctuation and symmetry breaking.   However for accurate and stable computation we had to keep the time step size in the order of $10^{-7}$ or less due to the restriction by the CFL condition.   Integration in terms of such time step requires too long computation time to attain relevant amount of result.   Therefore we do not pursue this issue any more in this work.

Fig. 20 shows time dependence of flowrate driven by such inlet boundary traction of Fig. 17.   First of all, one can notice discontinuous large peaks of flowrate (instant increase and decrease of flowrate) at those moments when step increase of traction is applied.   For example, at the moment of enforcing $\Delta p=300$ to the liquid at rest, i.e., at $t=0$ the flowrate reaches as high as 5860 and then very rapidly (but continuously or densely in more correct mathematical term) decreases to 2.73.   This is an artifact that originates from the creeping flow assumption.   The magnitude of instantaneous jump corresponds to the steady flowrate instantly attained by the inertialess Newtonian liquid of viscosity $s\eta$ under the same step traction.    The governing equations in the inertialess approach do not contain time derivative of velocity, and on the contrary **c** or **h** always exhibits continuous variation.    Thus such kind of temporally discontinuous velocity does not give rise to any computational problem.    Certainly we can recover the continuity of velocity if we include in the formulation some finite density however small it is.    Even though from Fig. 20 it is not clear, the flowrate exhibits oscillation starting from $t$=1.1, which has been more explicitly demonstrated in Figs. 19 and 21.    We now delve into these time periodical phenomena.

In Fig. 21, the flowrate variations in the time interval of 0.3 for $\Delta p$=360, 400 are depicted together with their FFT spectra.    One can examine almost (but not perfect yet) periodical oscillation, and the first large peak especially in FFT spectrum of Fig. 21 (b) reflects immaturity of periodicity.    By the same reason brought out for the curve in Fig. 19(b), one period in flowrate corresponds to a half cycle of general flow field fluctuation in Fig. 21 (a).    Thus in reality, $\theta f_1$ the location of the first characteristic peak in FFT spectra is twice the actual first characteristic frequency ($\theta f_1/2$).    In other words, on the spectral curve of Fig. 21 (a) only even multiples of the characteristic frequency appear due to the equivalence expressed by alternating half cycles in mirror image symmetry.    On the other hand, at the traction $\Delta p$=400 such symmetry of alternating half cycles is broken by the introduction of strong odd multiple modes (modes at $\theta f_1/2$, $3\theta f_1/2$, $5\theta f_1/2$ in the FFT spectrum of Fig. 21b).    If we compare even multiple modes only (modes at $\theta f_1$, $2\theta f_1$, $3\theta f_1$) between (a) and (b), we notice increase of characteristic frequency with increasing traction.    However at $\Delta p$=400 the amplitude of odd multiple modes becomes comparable to or even larger than that of even ones.    In the inset of Fig. 20, one may observe that such odd multiple harmonics are introduced into this elastic flow instability around $t\approx3.8$, where the region is marked with red circle.    We regard this as another symmetry breaking in flow dynamics, since the equivalence of alternating half cycles disappears.

Figs. 22 and 23 demonstrate streamlines near contraction to illustrate their temporal variation in the flow domain at specified instants under $\Delta p$=360 and 400, respectively.    The time instants are also marked with red dots on the curves of flowrate in Figs. 21 (b) and (c).    To help clear observation, vertical or horizontal red dotted lines are added in the figures.    In both Figs. 22 and 23, fluctuating streamlines are expressed as alternating height of right and left vortices or their centers, and skewness of center streamlines.    Also the streamlines at $t$=3.465 and 3.4818 in Fig. 22 suggest one evidence of mirror image symmetry of alternating half cycles under $\Delta p$=360.    On the other hand, in this flow regime of the Boger-type liquid ($n$=2) we have not witnessed such flow instability along the downstream channel wall displayed in Fig. 16(b).

# 5. Discussion and conclusion

In this work, comparing with conventional methods IE1 and CN2, we briefly examined new time integration algorithm DC2 in 4:1 planar contraction flows. The approximation procedure suggested here is a mixed method for the equation of motion (implicit in velocity and explicit in $\mathbf{c}$ or $\mathbf{h}$) and implicit for the constitutive equation. In applying this numerical tool, larger portion of computation time is assigned to solving Step (i) in Eqs. (A4) and (A5) since in that step we have more unknowns such as nodal variables for velocity, pressure and DEVSS variables. In addition to the speed and applicability in the whole range of $s$, this decoupling procedure endows another advantage in the multi-modal approach. In modeling real polymeric fluid flow in order to improve the description of material property, multi-Maxwellian mode is applied [2,3] almost without exception, where we are given with multiple sets of constitutive equations. Since with this decoupled algorithm each set can be numerically solved independently, the computation time only arithmetically increases with the number of modes for solving constitutive equations.

The tensor-log formulation implemented in this study carries one drawback, the mathematical complexity in transforming into $\mathbf{h}$-formulation, which hinders establishing closed form equations for most equations. On the other hand, with the Leonov model the mathematical form in $\mathbf{h}$-formulation is the simplest of all viscoelastic models due to $\det \mathbf{c} = 1$ or $\operatorname{tr} \mathbf{h} = 0$. Therefore at least in 2D we can obtain the closed form (9), which enables easy application of new temporal marching algorithm as well as direct computation of steady flow. Steady solutions obtained in such a way are proven to be free from the numerical degeneration according to mesh refinement.

In 2D contraction problem we considered only creeping flows in order to investigate purely elastic phenomena in highly nonlinear regime. The solution for pressure difference as high as 50 and higher with highly shear thinning fluid ($n$=0.1) exhibits severe fluctuation with low temporal correlation. The unstable flow has been displayed as variation of corner vortices, fluctuation of flowrate and propagation of alternating gradients of field variables along the downstream wall. It has been verified that to describe this elastic instability one requires spatial discretization refined enough to resolve the structural detail of such phenomena. The solution in the full flow domain reveals severe symmetry breaking in the narrow downstream channel, while the rheological variables including corner vortices remain almost symmetric in the domain before contraction.

We can observe elastic instability drastically distinct from that of the shear thinning liquid when the Boger-type fluid ($n$=2) is assumed. The unstable flow dynamics becomes periodic for both half and full domain problems. Flow asymmetry expressed as alternating corner vortices occurs in the upstream reservoir in the full domain. By the comparison of half and full domain solutions, we conclude that symmetry breaking attenuates the intensity of instability by lessening the amplitude of flowrate and approximately doubling the frequency of its flowrate oscillation. Such asymmetry is manifested as equivalently repeating half cycles in the exact mirror image, that is, only the even multiples of characteristic frequency exist in the FFT spectra. However, the mirror image symmetry between alternating half cycles disappear when $\Delta p \geq 400$, even if the dynamics still remains perfectly periodic. At this highly nonlinear flow regime, odd multiple modes of the characteristic harmonic play rather significant role. On the other hand along the narrow downstream channel, no propagating concentrated gradients of variables have been expressed with this liquid type.

As an additional remark, in this work we have never observed such back-shedding of corner vortices reported in the paper [20]. We do not think such qualitative difference results from the difference between numerical methods employed. One possible reason may be that in this study the applied traction is still too low to describe such singular phenomena. However the current author thinks such discrepancy most probably originates from the fundamental dissimilarity existent in the constitutive equations implemented.

In real elastic flow, there exists highly nonlinear flow phenomenon called elastic turbulence [36]. It is considered quite useful for microfluidic devices where turbulence usually induced by high Reynolds number flow is impossible to attain [34,35]. That is, fluid elasticity is able to cause turbulent mixing in such small devices. Even if the elastic instability expressed here may share the same origin with theses real phenomena, correct description of elastic turbulence necessitates full 3D modeling. In 2D modeling, the streamlines never intersect, and thus mixing effect cannot become possible at all. Turbulence is certainly a full 3D phenomenon, and therefore full geometry modeling definitely with removal of symmetric boundary is essential to explain this unique flow effect. Actually the unstable flow phenomena explained here cannot exactly coincide with real flow instability due to the approximation of 2D flow domain. In this 2D flow setting, we presume infinite thickness of the channel in the direction normal to domains in Fig.1. Thus, for example, the vortex oscillation expressed by the solution implies the exactly same behavior in every cross-section of this infinitely thick channel and thus it seems impossible to realize. On the contrary the fluctuation of vortices would preferably appear as traveling bundle-like streams observed in [37]. However the computational procedure and its solution provided in this work show us some possibility of successfully modeling elastic instability or elastic turbulence in 3D setting.

**Acknowledgements**

This research was supported by Basic Science Research Program through the National Research Foundation of Korea (NRF) funded by the Ministry of Education, Science and Technology (2010-0009654).   The author also acknowledges the financial support from the Institute for Chemical Research in Kyoto University for his sabbatical stay and is very grateful to Prof. Hiroshi Watanabe for helpful discussions and care during the stay.

# Appendix A. Conventional and decoupled time integration schemes of order 1 and 2

Before introducing decoupled discrete temporal advancing scheme, we briefly discuss the conventional procedure of numerical time integration such as implicit Euler and Crank-Nicolson methods, which are employed for comparison with the algorithm developed in this study.

The set of Eqs. (1) and (7) may be rewritten in the following generic form:

$$Re\frac{\partial v_i}{\partial t} = \Lambda_i\left(v_l, c_{mn}, I_1, p\right), \qquad \frac{\partial v_1}{\partial x} + \frac{\partial v_2}{\partial y} = 0, \qquad \frac{\partial h_{ij}}{\partial t} = \Psi_{ij}\left(v_l, h_{mn}\right). \tag{A1}$$

Here $\Lambda_i$ and $\Psi_{ij}$ include all the terms in Eqs. (1) and (7) except the time derivative. The $h_{ij}$ variables can be easily converted to $c_{ij}$ in terms of the relation (8) and then applied to the momentum equation. The stress relation in Eqs. (2) contains $\left(I_1\right)^q$. Thus in the equation of motion $I_1 = c_{ii}$ is explicitly but redundantly included in the list of independent variables for $\Lambda_i$, since the new 2nd order integration scheme adopts its approximation different from that of $c_{mn}$. The standard numerical integration procedure such as the implicit Euler (IE1) and the Crank-Nicolson (CN2) methods may be expressed as

**IE1**:

$$Re\left(v_i^{k+1} - v_i^k\right) = \Delta t \cdot \Lambda_i\left(v_l^{k+1}, c_{mn}^{k+1}, I_1^{k+1}, p^{k+1}\right), \qquad \frac{\partial v_1^{k+1}}{\partial x} + \frac{\partial v_2^{k+1}}{\partial y} = 0,$$

$$c_{ij}^{k+1} - c_{ij}^k = \Delta t \cdot \Phi_{ij}\left(v_l^{k+1}, c_{mn}^{k+1}\right) \quad \text{or} \quad h_{ij}^{k+1} - h_{ij}^k = \Delta t \cdot \Psi_{ij}\left(v_l^{k+1}, h_{mn}^{k+1}\right). \tag{A2}$$

**CN2**:

$$Re\left(v_i^{k+1} - v_i^k\right) = \frac{\Delta t}{2} \cdot \left[\Lambda_i\left(v_l^{k+1}, c_{mn}^{k+1}, I_1^{k+1}, p^{k+1}\right) + \Lambda_i\left(v_l^k, c_{mn}^k, I_1^k, p^k\right)\right], \qquad \frac{\partial v_1^{k+1}}{\partial x} + \frac{\partial v_2^{k+1}}{\partial y} = 0,$$

$$c_{ij}^{k+1} - c_{ij}^k = \frac{\Delta t}{2} \cdot \left[\Phi_{ij}\left(v_l^{k+1}, c_{mn}^{k+1}\right) + \Phi_{ij}\left(v_l^k, c_{mn}^k\right)\right] \quad \text{or}$$

$$h_{ij}^{k+1} - h_{ij}^k = \frac{\Delta t}{2} \cdot \left[\Psi_{ij}\left(v_l^{k+1}, h_{mn}^{k+1}\right) + \Psi_{ij}\left(v_l^k, h_{mn}^k\right)\right]. \tag{A3}$$

Here $\Delta t$ denotes the increment in one time step and the superscript designates the number of the step. As is well known, the implicit Euler and Crank-Nicolson integration algorithms endow the 1st and 2nd order accuracy, respectively.

Just for the enhancement of computation speed, we devise an algorithm that solves the constitutive equations separately from the total set by way of simple decoupling. In the 1st order method, backward approximation for velocity but forward for **c**-tensor is employed for the momentum balance, and then

$$Re\left(v_i^{k+1} - v_i^k\right) = \Delta t \cdot \Lambda_i\left(v_l^{k+1}, c_{mn}^k, I_1^k, p^{k+1}\right)$$ results, which is still 1st order convergent. As a result solving the equation of motion together with the continuity equation yields the value $v_i^{k+1}$ and its substitution into the discrete constitutive equation enables solution for $c_{ij}^{k+1}$ or $h_{ij}^{k+1}$ independently.

However this naïve modification may incur some singular artifact in the critical situation of highly elastic creeping flow. In viscoelastic flow dynamics, due to high viscosity the inertial term often becomes negligible even at high $De$. While specification of large value for the retardation parameter $s$ dramatically stabilizes the computational procedure, it may at the same time spoil description of elastic instability as well as liquid property. Thus the modeling of high $De$ flow with small retardation parameter in creeping flow becomes often favorable to get the insight into various elastic effects. In this study the main concern is to develop a computational tool that can describe high $De$ flow in the whole range of the retardation parameter $s$ ($0 \leq s \leq 1$).

In order to illustrate the problem possibly existent in the above simple decoupling idea for small retardation, we consider the extreme case of $s=0$. In the creeping flow the equation of motion in Eqs. (A1) simplifies to $\Lambda_i\left(v_l, c_{mn}, I_1, p\right) = 0$, and the condition of $s=0$ again removes the velocity variable in $\Lambda_i$. Thus the momentum equation in this simple decoupled scheme does not generate velocity update $v_i^{k+1}$. Under pressure difference between inlet and outlet of the flow channel, setting the vanishing $s$ in creeping flow incurs minor numerical problem that the initial velocity reaches infinity due to instantaneous elastic response. Accordingly we do not entirely neglect the retardation term in numerical computing. However one has to expect error increase via numerical approximation with $s$ approaching 0 in this simple decoupled algorithm.

In order to remedy this shortcoming while maintaining the advantage endowed by decoupling, we suggest the following modification as the 1st (DC1) and 2nd (DC2) order decoupled procedures:

**DC1**

Step (i)   $Re\!\left(v_i^{k+1} - v_i^k\right) = \Delta t \cdot \Lambda_i\!\left(v_l^{k+1}, c_{mn}^{**}, I_1^k, p^{k+1}\right),$   $\dfrac{\partial v_1^{k+1}}{\partial x} + \dfrac{\partial v_2^{k+1}}{\partial y} = 0,$

where   $c_{mn}^{**} - c_{ij}^k = \Delta t \cdot \Phi_{ij}\!\left(v_l^{k+1}, c_{mn}^k\right),$

Step (ii)   $h_{ij}^{k+1} - h_{ij}^k = \Delta t \cdot \Psi_{ij}\!\left(v_l^{k+1}, h_{mn}^{k+1}\right).$                    (A4)

**DC2**

Step (a)   $h_{ij}^* - h_{ij}^k = \Delta t \cdot \Psi_{ij}\!\left(v_l^k, h_{mn}^k\right)$ ➔ obtain $c_{ij}^*$ or $I_1^* = c_{ii}^*,$

Step (i)   $Re\!\left(v_i^{k+1} - v_i^k\right) = \dfrac{\Delta t}{2} \cdot \left[\Lambda_i\!\left(v_l^{k+1}, c_{mn}^{**}, I_1^*, p^{k+1}\right) + \Lambda_i\!\left(v_l^k, c_{mn}^k, I_1^k, p^k\right)\right],$

$\dfrac{\partial v_1^{k+1}}{\partial x} + \dfrac{\partial v_2^{k+1}}{\partial y} = 0,$

where   $c_{mn}^{**} - c_{ij}^k = \Delta t \cdot \Phi_{ij}\!\left(v_l^{k+1}, c_{mn}^k\right),$

Step (ii)   $h_{ij}^{k+1} - h_{ij}^k = \dfrac{\Delta t}{2} \cdot \left[\Psi_{ij}\!\left(v_l^{k+1}, h_{mn}^{k+1}\right) + \Psi_{ij}\!\left(v_l^k, h_{mn}^k\right)\right].$       (A5)

In both the methods, the main idea is as follows.   In order for the equation of motion to contain velocity variables explicitly even for $Re = s = 0$, we substitute the forward (in **c**) and backward (in **v**) approximation of the constitutive equation for **c** of the equation of motion, which is denoted as $c_{mn}^{**}$.   However for $\left(I_1\right)^n$ in DC1 we still employ $\left(I_1^k\right)^n$ in the equation of motion for computational simplicity. After completing the velocity update $v_i^{k+1}$, the procedure then determines $h_{ij}^{k+1}$ or $c_{ij}^{k+1}$ in Step (ii) with velocity known.

The 2nd order decoupled method DC2 is mainly based on the Crank-Nicolson scheme.   Even though the auxiliary Step (a) required for obtaining $I_1^* = c_{ii}^*$ and the Euler approximation for $c_{mn}^{**}$ are only the 1st order accurate, existence of the factor $\Delta t$ in front of $\Lambda_i$ in the momentum equation automatically enforces the 2nd order convergence.   Also note that except for Step (ii) all the discrete equations are linear in their unknowns when $Re$=0.   Thus for example, Newton's iteration implemented here for solving nonlinear systems is necessary only for Step (ii) in creeping flow. Moreover further linearization can also be made for Step (ii) as follows. In $\Psi_{ij}$ except for the terms $v_1 \dfrac{\partial h_{ij}}{\partial x} + v_2 \dfrac{\partial h_{ij}}{\partial y}$ we employ the predetermined approximation $h_{ij}^*$ for $h_{ij}^{k+1}$ and then the equation of Step (ii) again becomes linear. Since the introduction of this additional approximation has not incurred noticeable difference at least for the results presented in this work, no distinction is made between these algorithms.

The only difference of the method DC2 from the scheme suggested by D'Avino and Hulsen [21] is the implementation of preliminary Step (a) introduced due to additional nonlinearity in the stress relation of the Leonov model. Thus in this paper, we do not thoroughly investigate the characteristics of the newly suggested time integration algorithms, even though some results of comparison with conventional methods (IE1 and CN2) are presented here for the contraction flow. Its validity with the Oldroyd-B and Giesekus equations has already been examined [21].

Tabel 1. Mesh details for half and full domain problems, limit pressure difference and corresponding Deborah number in steady state of flow.

| | Mesh 1 | Mesh 2 | Mesh 3 | Mesh 4 | Mesh 5 | Mesh 6 | Mesh 7 | Mesh 8 |
|---|---|---|---|---|---|---|---|---|
| **Smallest element size (scaled by $H_0$)** | 0.18 | 0.12 | 0.1 | 0.08 | 0.05 | 0.035 | 0.02 | 0.01 |
| **Number of elements** (half domain) | 1400 | 3005 | 3361 | 4597 | 6929 | 11006 | 15683 | 30409 |
| **Number of unknowns** (half domain) | 15376 | 32068 | 35664 | 48492 | 72352 | 113908 | 161308 | 310392 |
| **Number of elements** (full domain) | 2800 | 6010 | 6722 | 9194 | 13858 | 22012 | 31366 | 60818 |
| **Number of unknowns** (full domain) | 29508 | 62364 | 69388 | 94684 | 142104 | 224748 | 319188 | ------- |
| **Limit traction force** (n=0.1) | ------- | ------- | 53.2 | ------- | 52.8 | 55.2 | 53.6 | 54.2 |
| **Limit Deborah number** (n=0.1) | ------- | ------- | 65.4 | ------- | 65.0 | 81.4 | 73.1 | 77.8 |
| **Limit traction force** (n=2) | 158 | 193 | 148 | 162 | 213 | ------- | ------- | ------- |
| **Limit Deborah number** (n=2) | 5.34 | 6.65 | 4.98 | 5.50 | 7.40 | ------- | ------- | ------- |

# List of Figures

the Leonov model under $\Delta p$=90 computed with Mesh 3.  Figure inset is an enlarged view near $t$=0.55.

Fig. 12. Snapshots of streamlines computed Mesh 3 near re-entrant corner in contraction flow under $\Delta p$=90 at different time instants (a) $t$=4.619, (b) 4.646, (c) 4.666, (d) 4.673.

Fig. 13. Snapshots of streamlines and elastic potential contour computed with Mesh 3 along the downstream channel in contraction flow under $\Delta p$=90 at different time instants (a) $t$=4.619, (b) 4.646, (c) 4.666, (d) 4.673.

Fig. 14. Temporal flowrate variation and its FFT spectrum of the data in the range of and $t$=3~5 ($n$=0.1) for (a) $\Delta p$=50 (Mesh 6) and (b) $\Delta p$=90 (Mesh 3).

Fig. 15. (a) Temporal flowrate variation and (b) its FFT spectrum of the data in the range of and $t$=1.5~3.5 ($n$=0.1) for $\Delta p$=80 for 2 different spatial discretizations (Mesh 1 and 3) of full flow domain.

Fig. 16. (a) Streamline near contraction at 2 time instants ($t$=3.024, 3.062) and (b) streamline and pressure contour along the downstream channel at 3 time instants ($t$=3.024, 3.062, 3.094) ($n$=0.1) for $\Delta p$=80 for Mesh 3 of full flow domain.

Fig. 17. Applied inlet boundary traction according to time for the full domain problem with the parameter $n$=2.

Fig. 18. Temporal flowrate variation of the full domain problem under $\Delta p$=300 for 2 different time range in (a) and (b), computed with 2 different time steps such as $\Delta t$=0.5×10$^{-5}$ ($\Delta t$ is fixed) and 1.25×10$^{-5}$ (in $t$=0~1.0895 $\Delta t$=2.5×10$^{-5}$ is applied).

Fig. 19. Temporal flowrate variation of the full and half domain problems under $\Delta p$=300 in the time range of (a) $t$=1.2~1.5 and (b) $t$=1.5~1.8 ($\Delta t$=0.5×10$^{-5}$).

Fig. 20. Temporal flowrate variation of the full domain problem under $\Delta p$ specified in Fig. 17.  Computation is performed with Mesh 1.  Figure inset is an enlarged view near $t$=3.8.

Fig. 21. Temporal flowrate variation and its FFT spectrum of the data in (a) $t$=3.2~3.5 with $\Delta p$=360, and (b) $t$=3.82~4.12 with $\Delta p$=400.

Fig. 22. Snapshots of streamlines computed with Mesh 1 near contraction corner

for $\Delta p$=360 at 3 time instants $t$=3.465 and 3.4818.

Fig. 23. Snapshots of streamlines computed with Mesh 1 near contraction corner for $\Delta p$=400 at 6 time instants $t$=4.0715, 4.0745, and 4.0835.

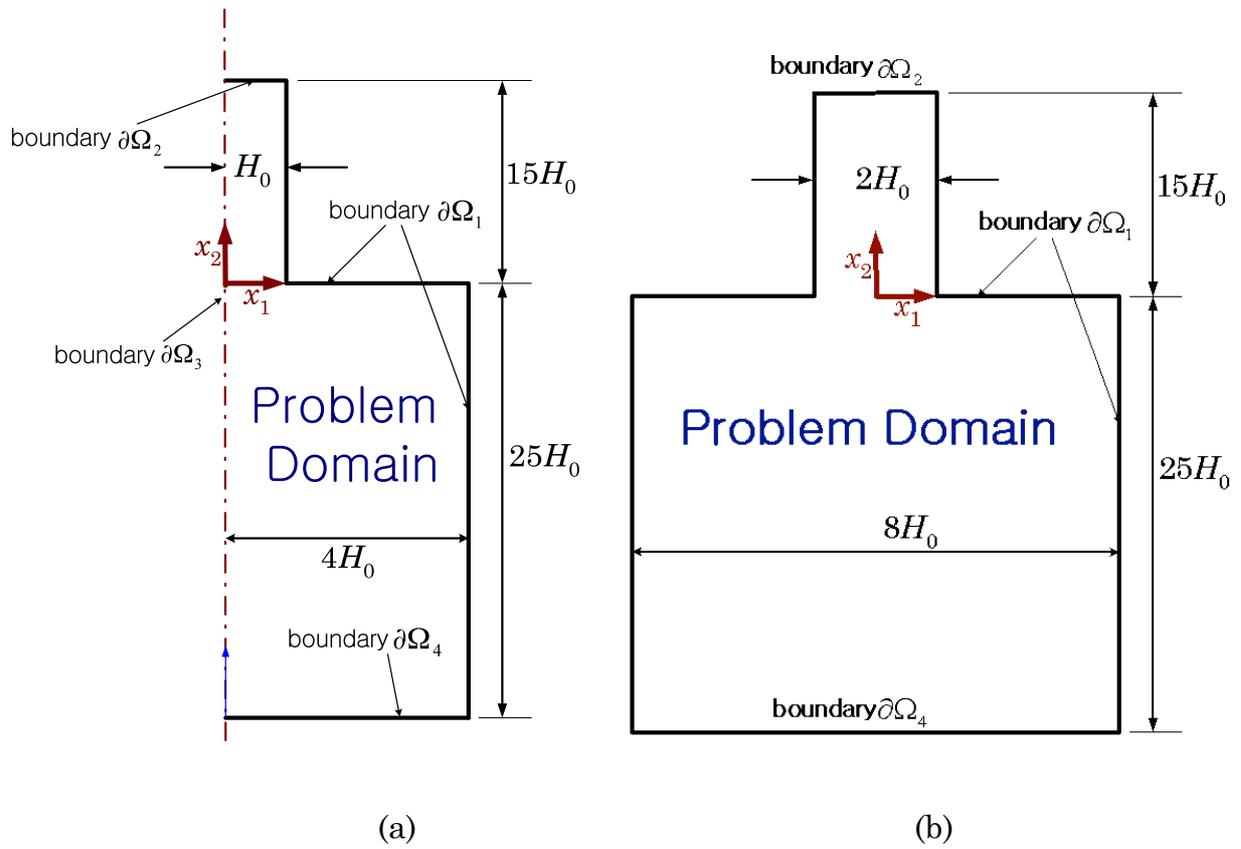

(a)                  (b)

$\partial\Omega_1$: no slip ($\mathbf{v=0}$)

$\partial\Omega_2$: traction free

$\partial\Omega_3$: symmetry

$\partial\Omega_4$: finite traction and stress free ($\mathbf{h=0}$)

Fig. 1

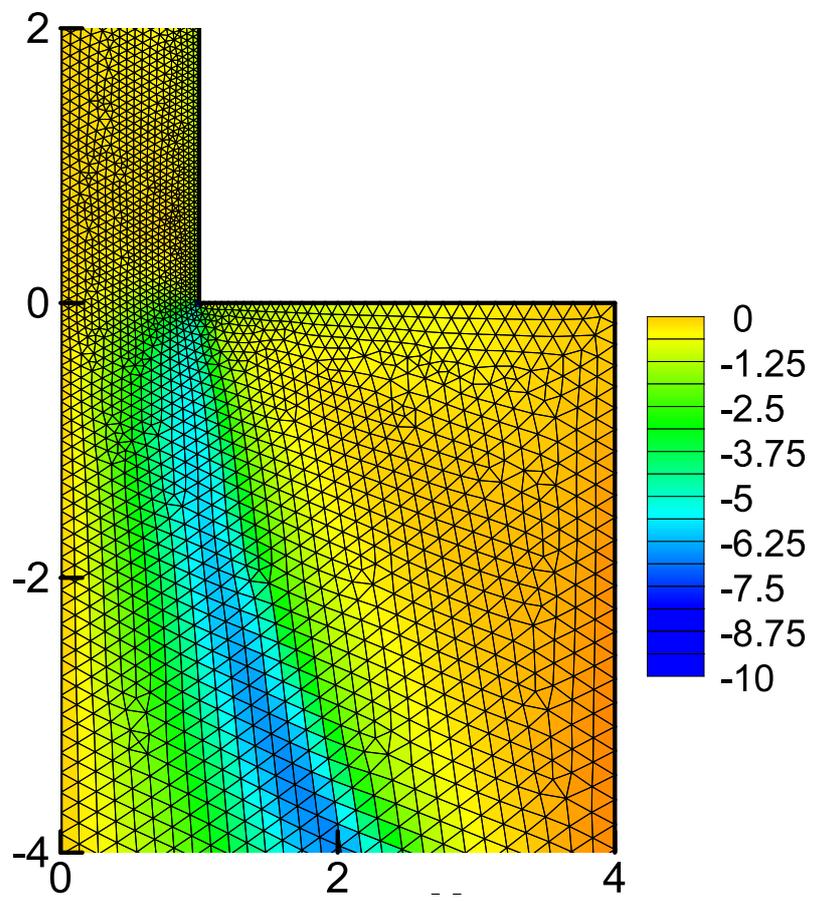

Fig. 2

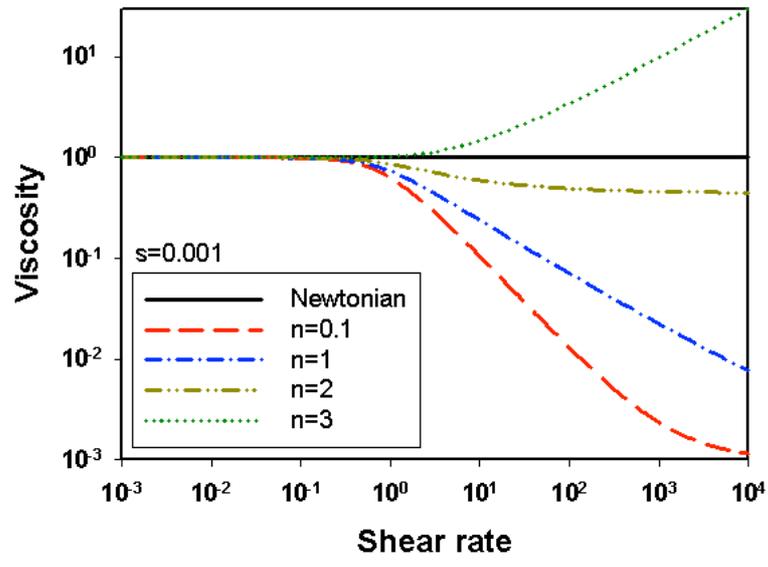

Fig. 3(a)

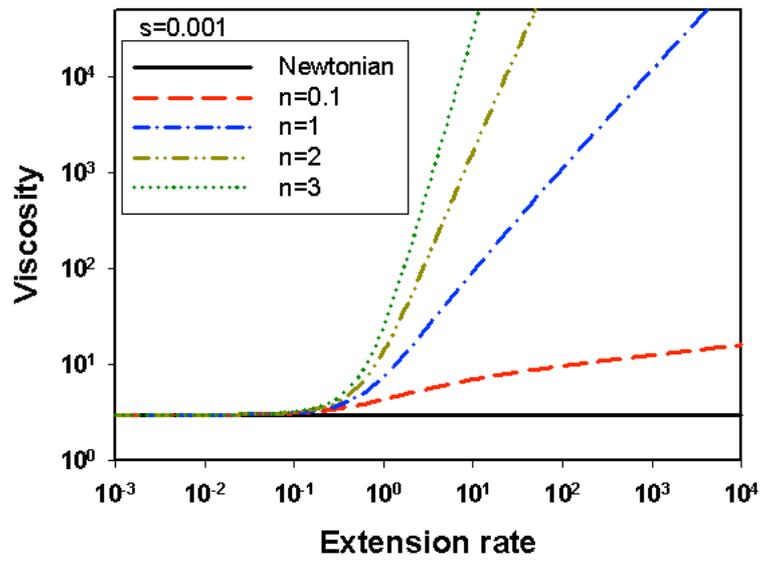

Fig. 3(b)

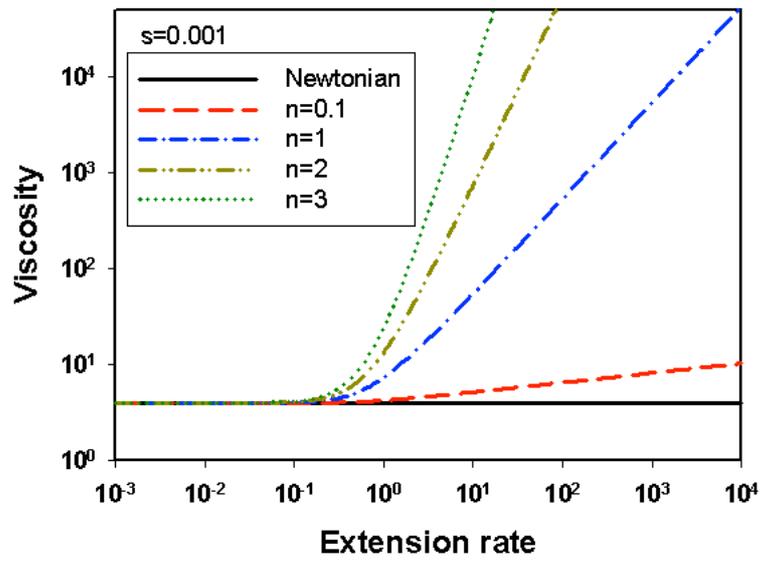

Fig. 3(c)

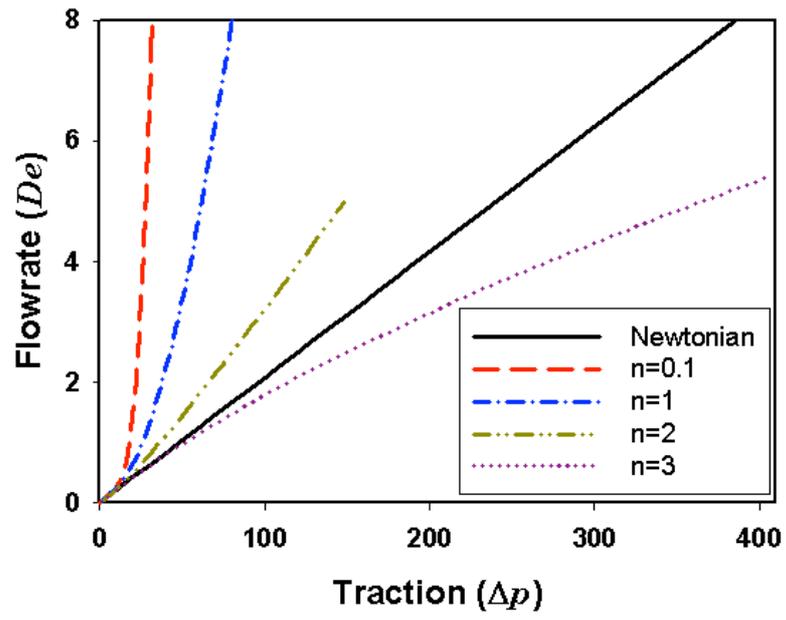

Fig. 4

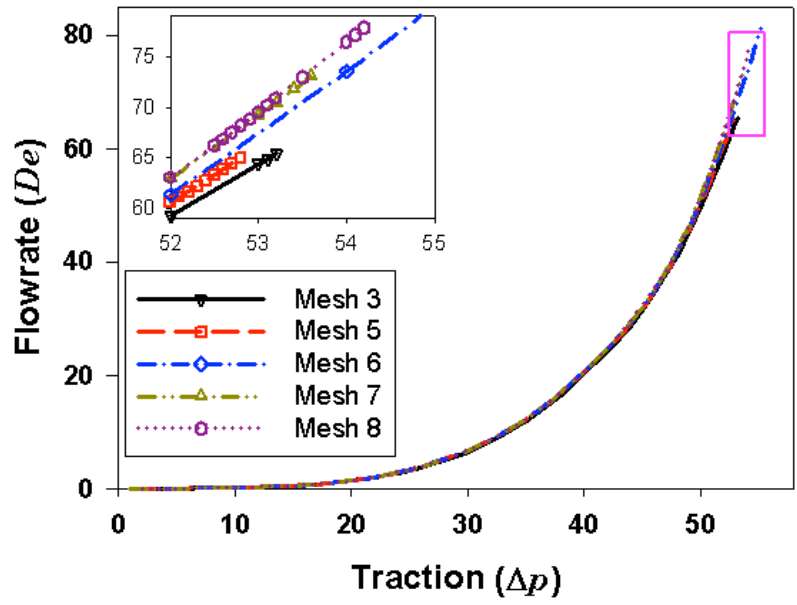

Fig. 5(a)

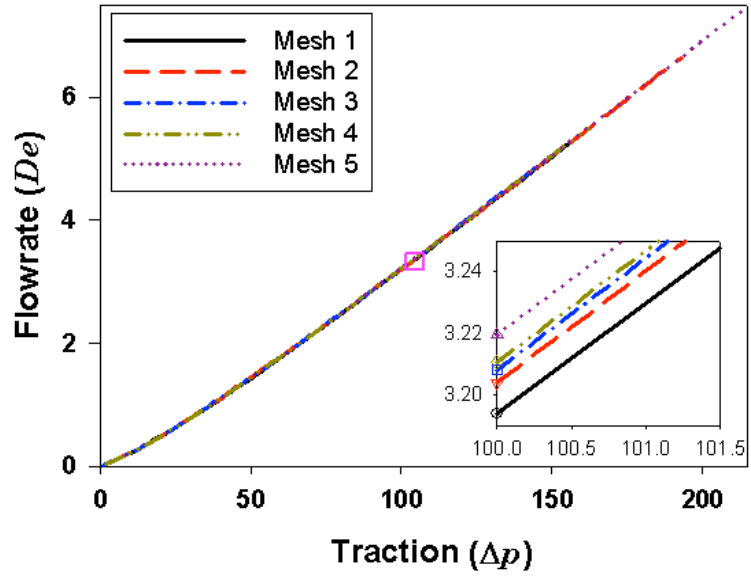

Fig. 5(b)

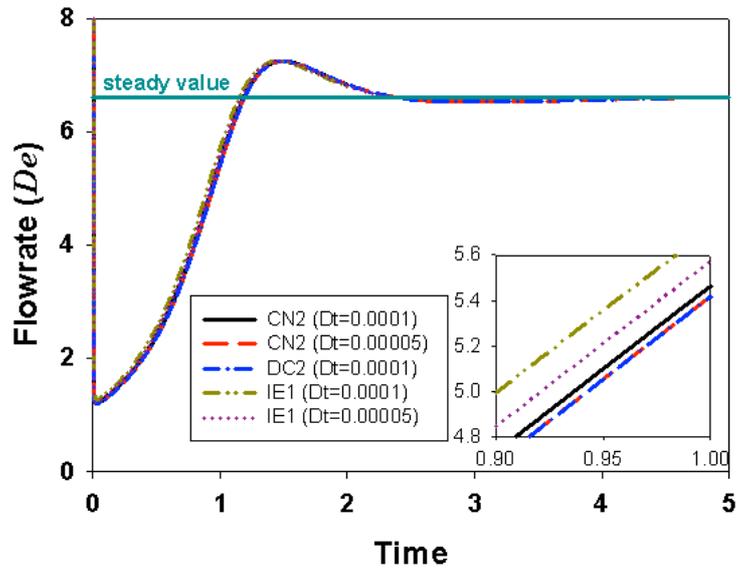

Fig. 6

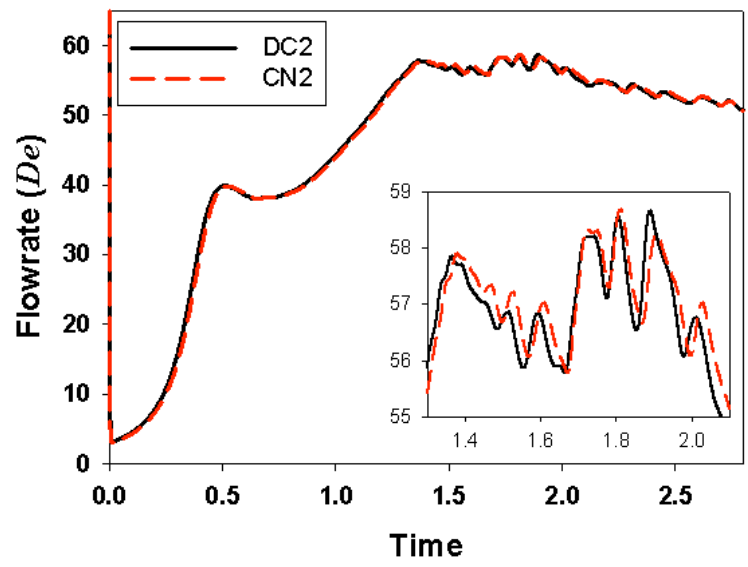

Fig. 7

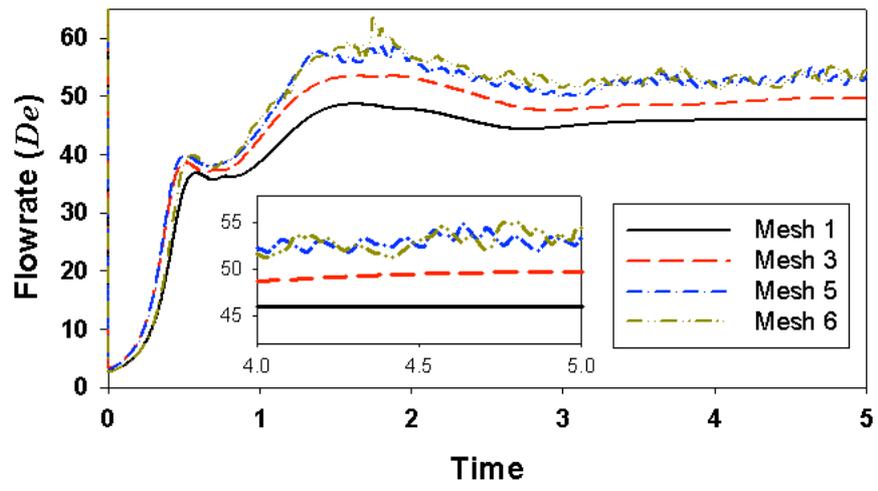

Fig. 8

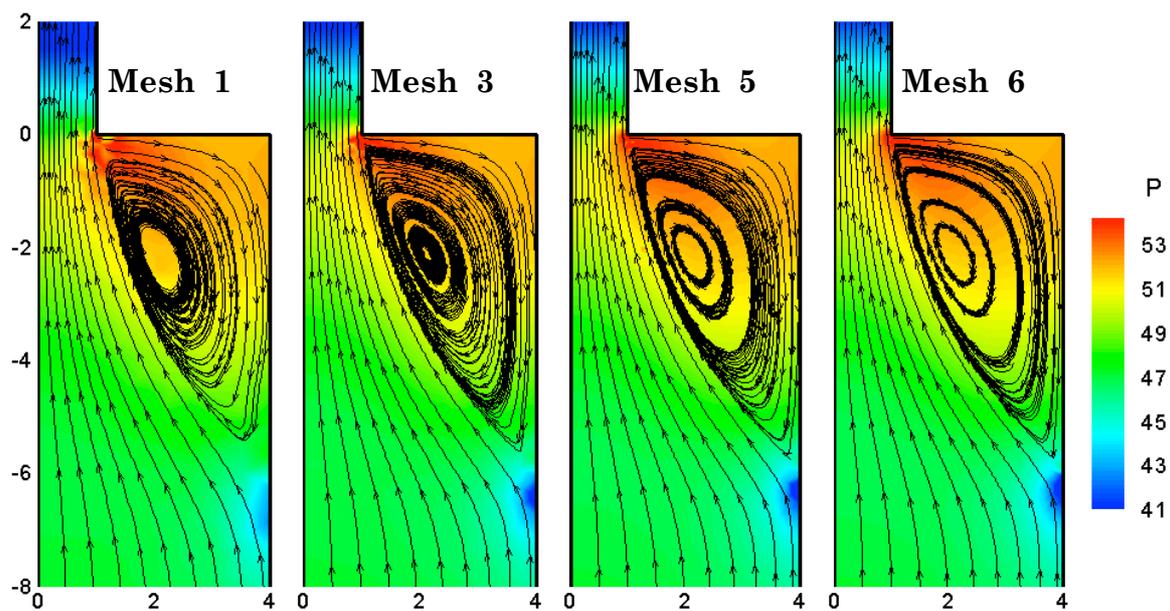

Fig. 9(a)

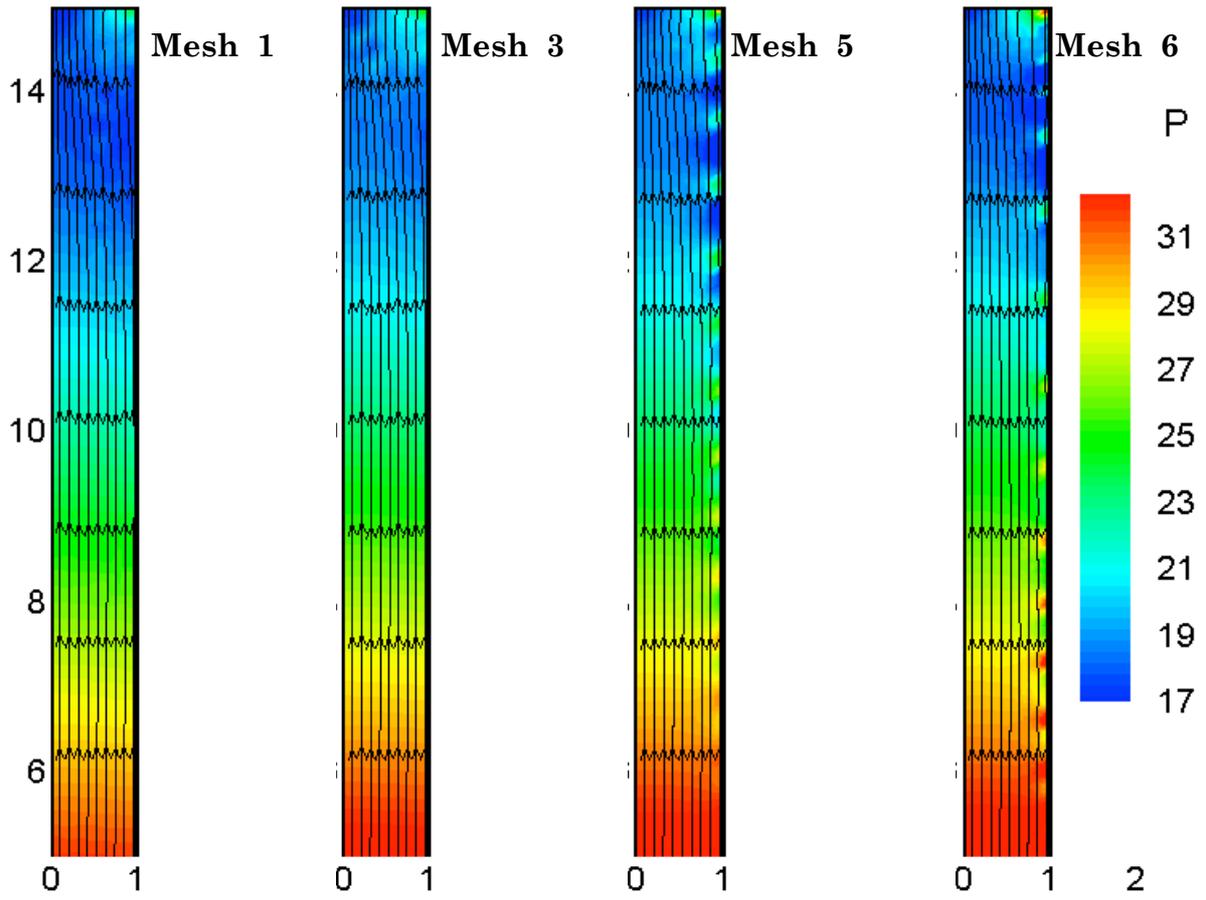

Fig. 9(b)

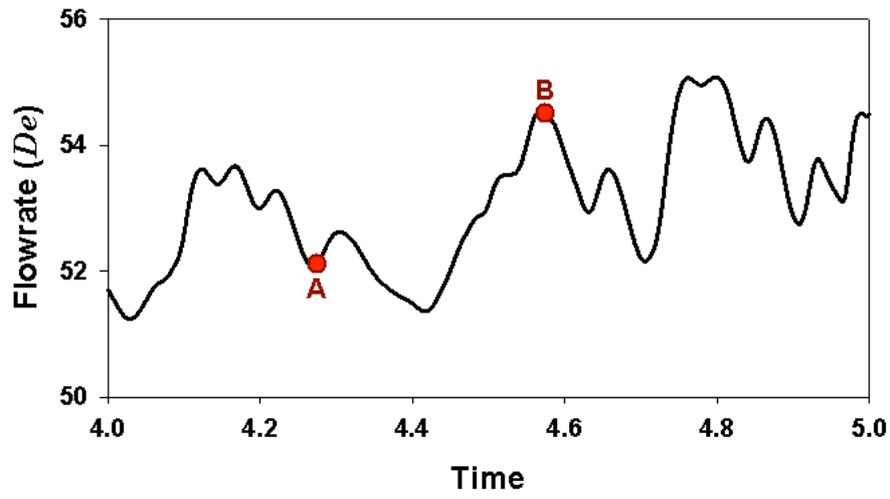

Fig. 10(a)

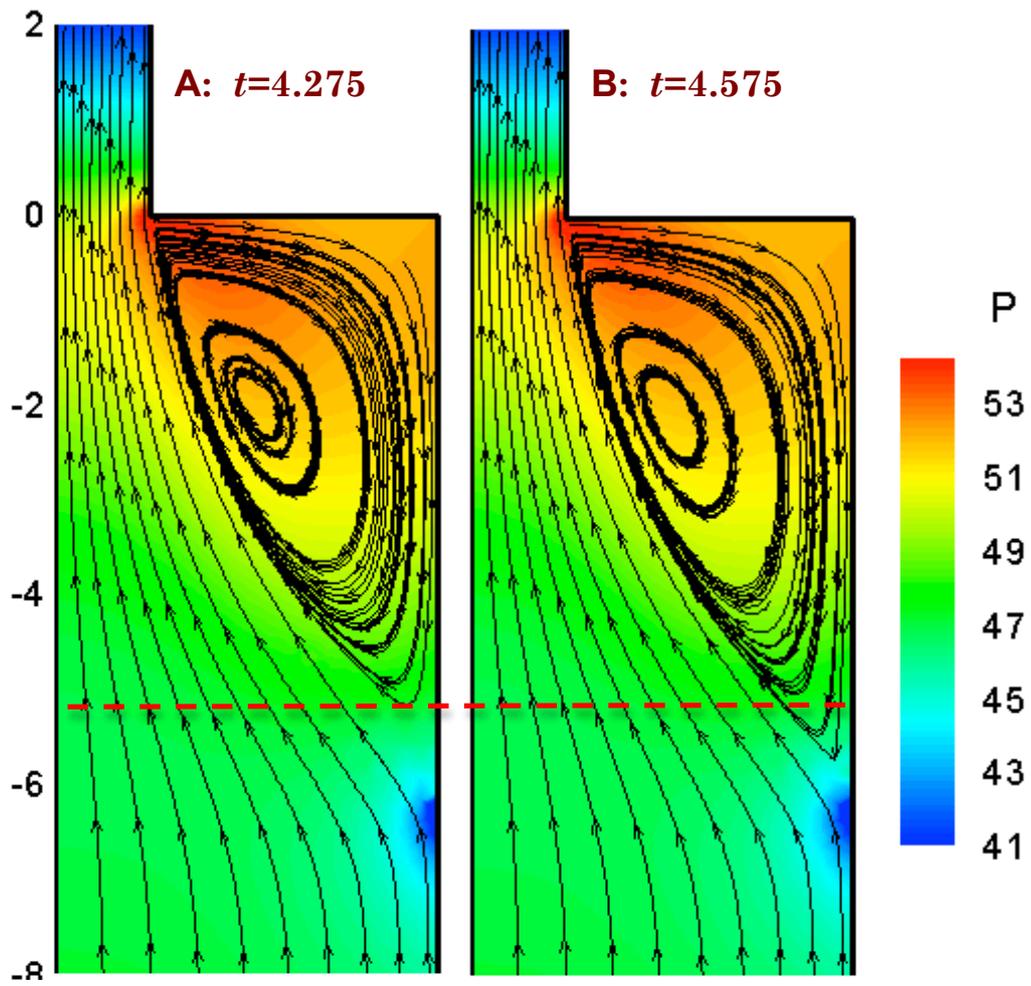

Fig. 10(b)

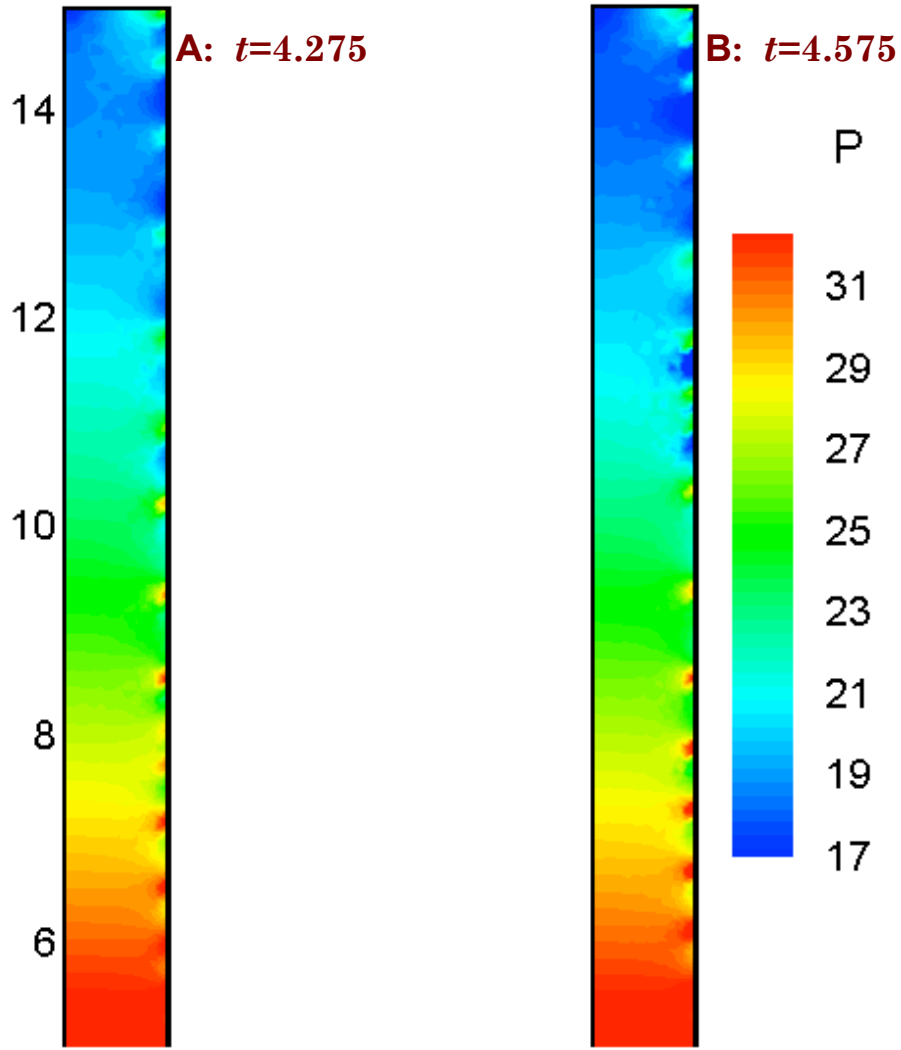

Fig. 10(c)

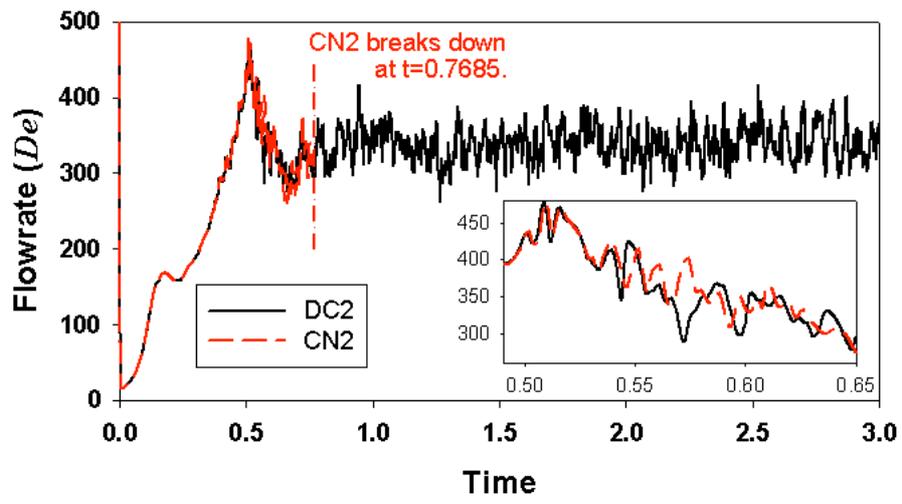

Fig. 11

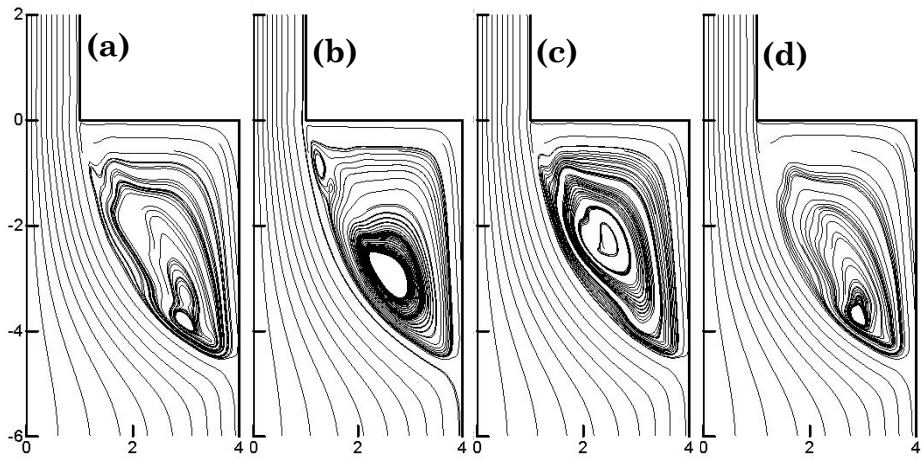

Fig. 12

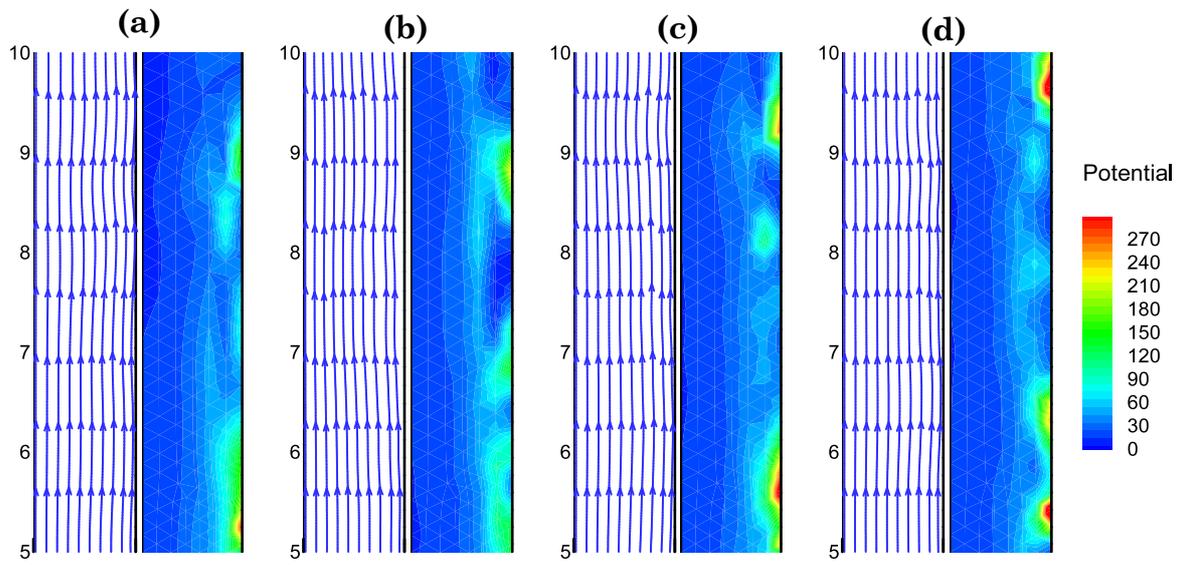

Fig. 13

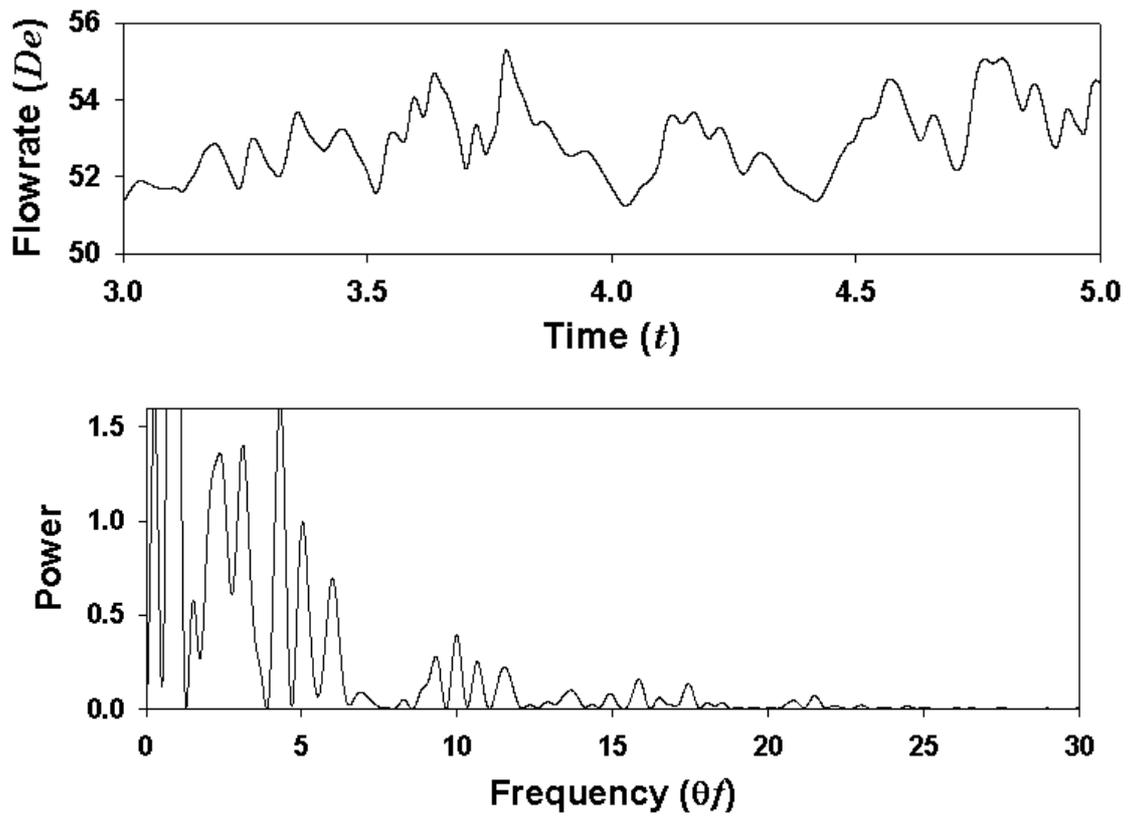

Fig. 14(a)

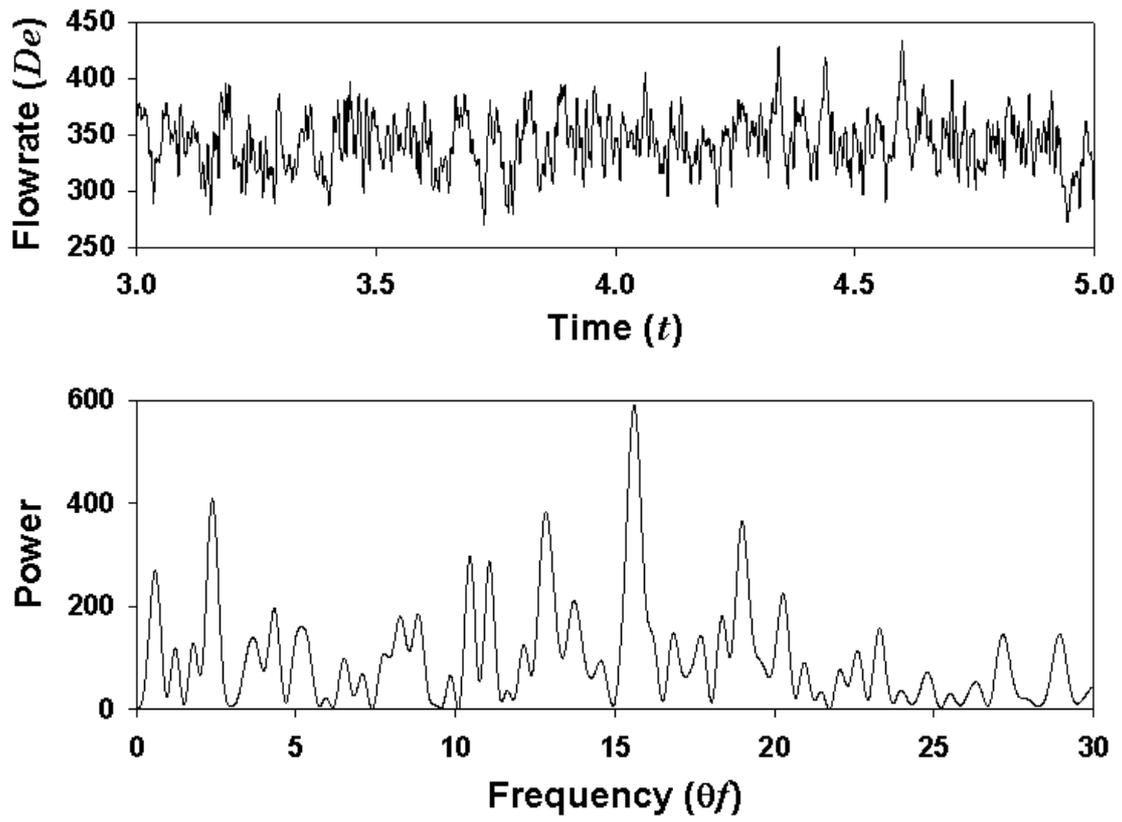

Fig. 14(b)

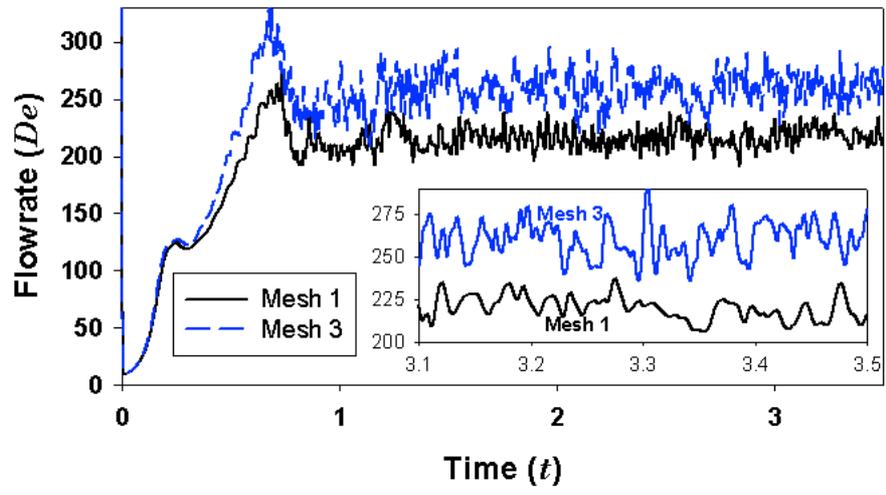

Fig. 15(a)

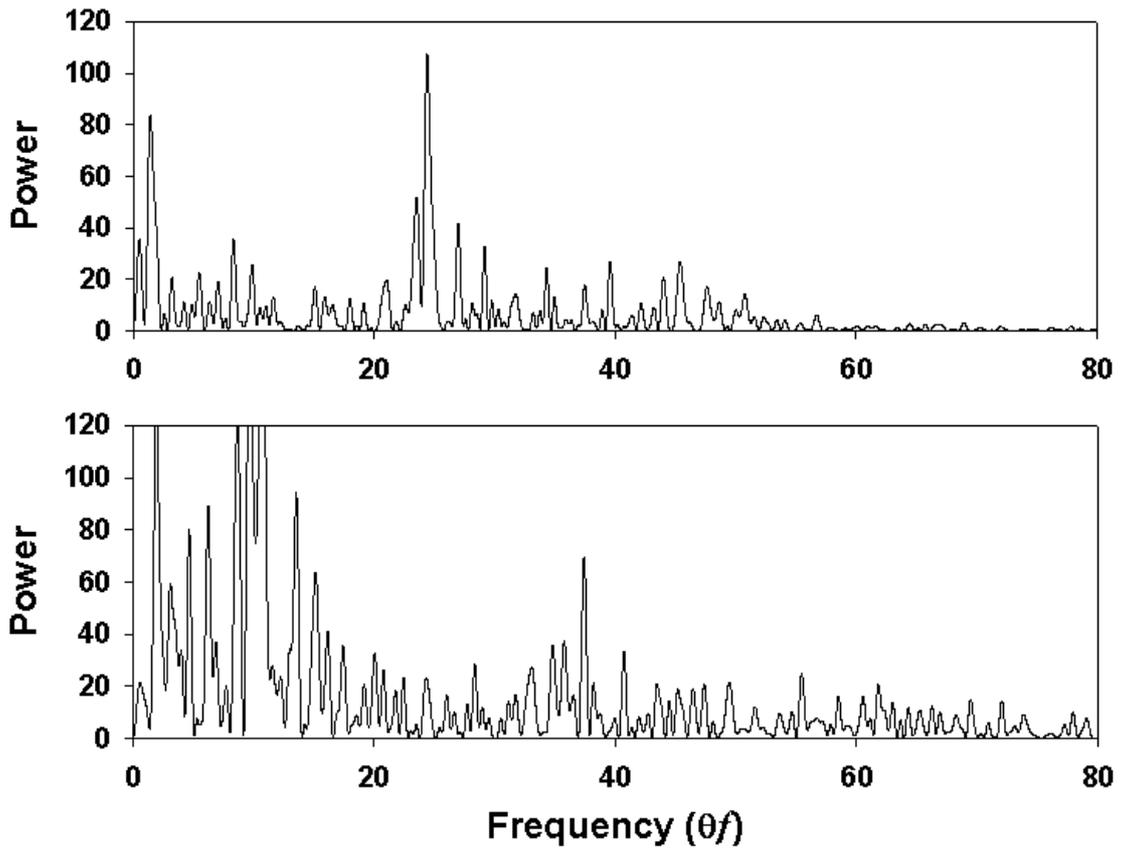

Fig. 15(b)

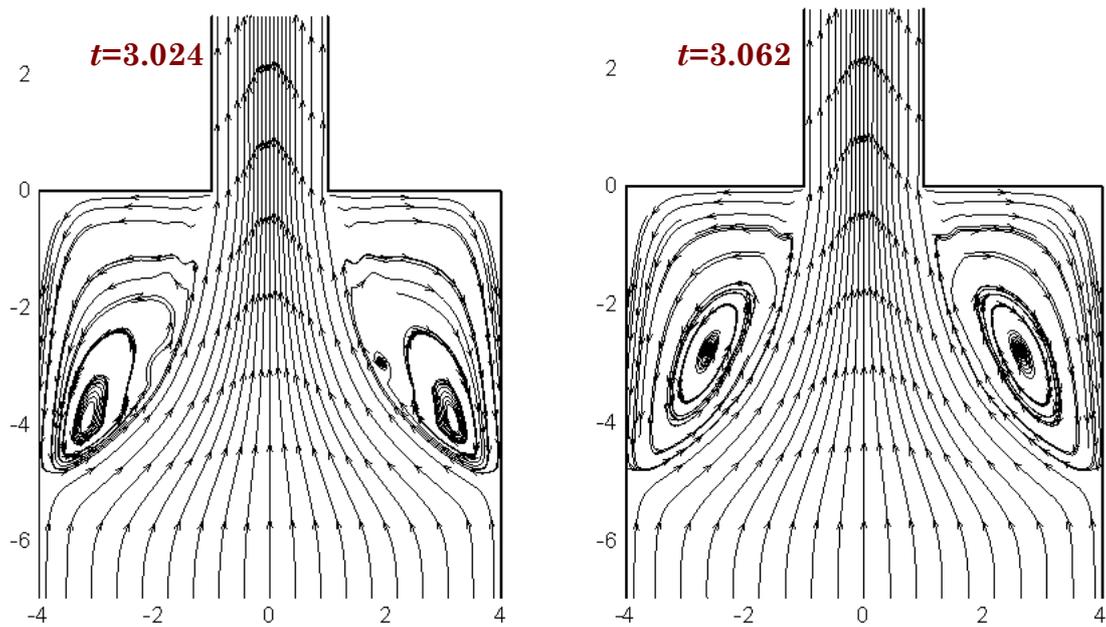

Fig. 16(a)

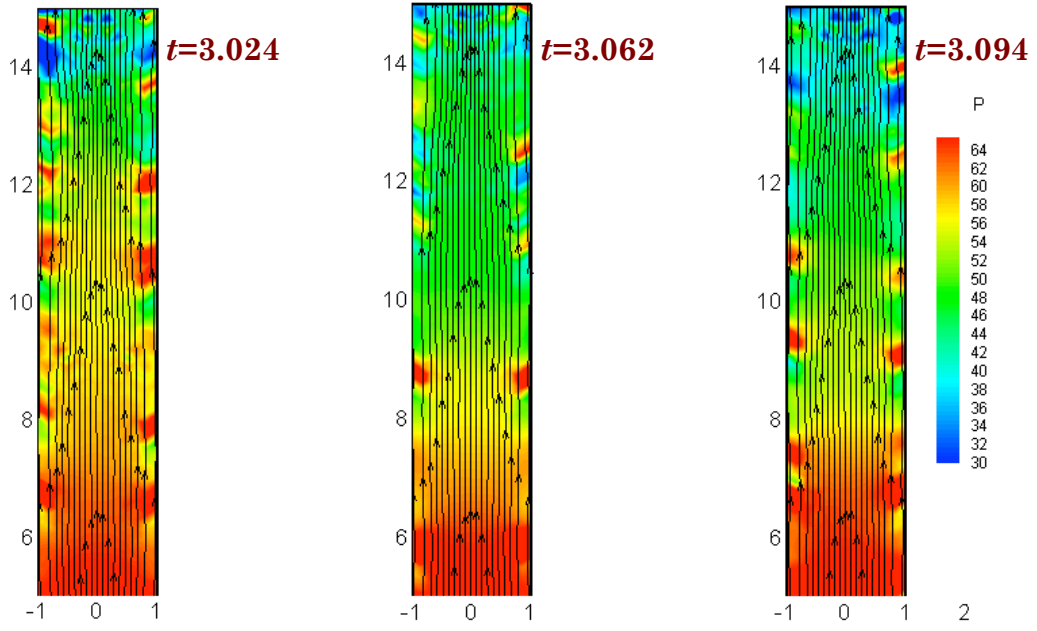

Fig. 16(b)

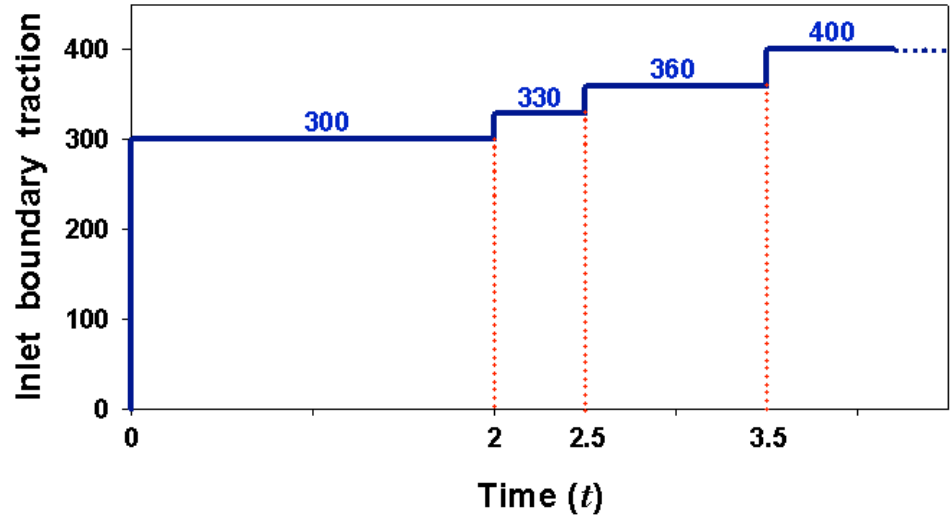

Fig. 17

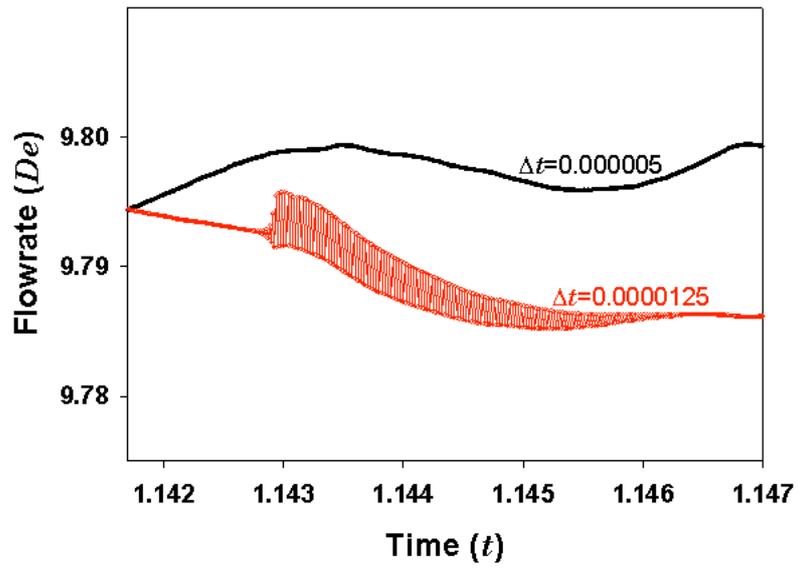

Fig. 18(a)

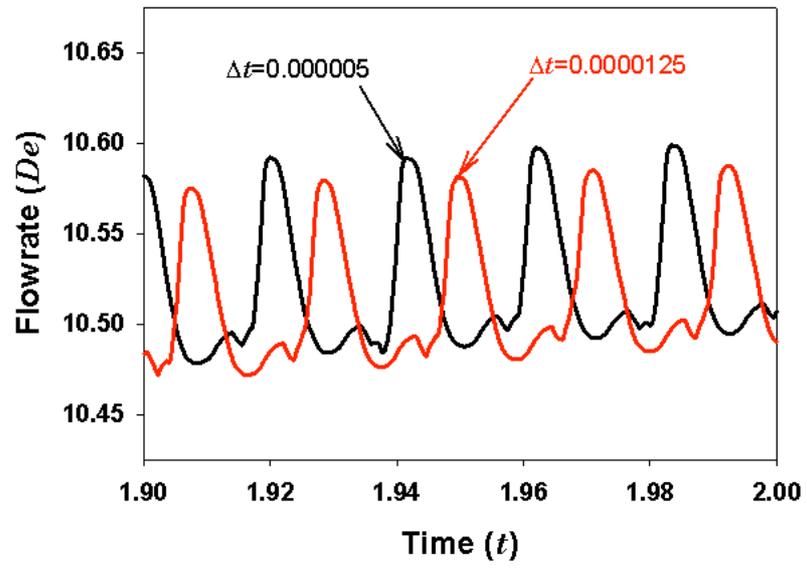

Fig. 18(b)

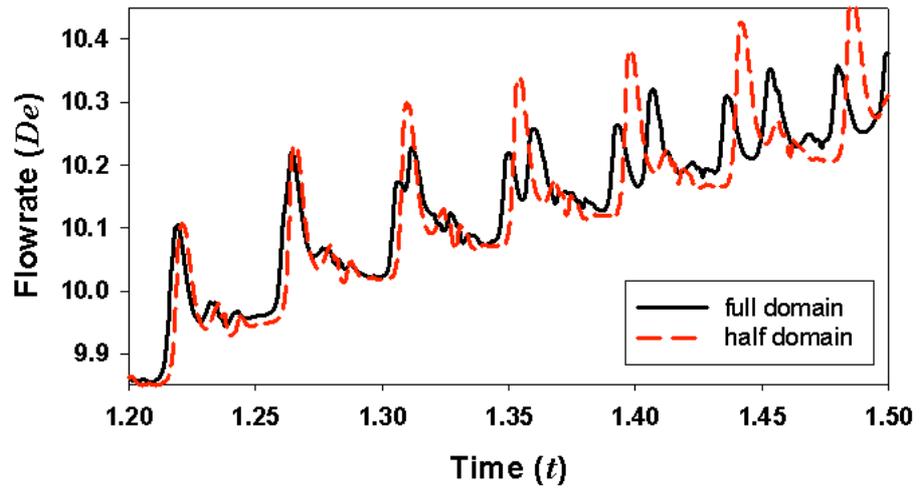

Fig. 19(a)

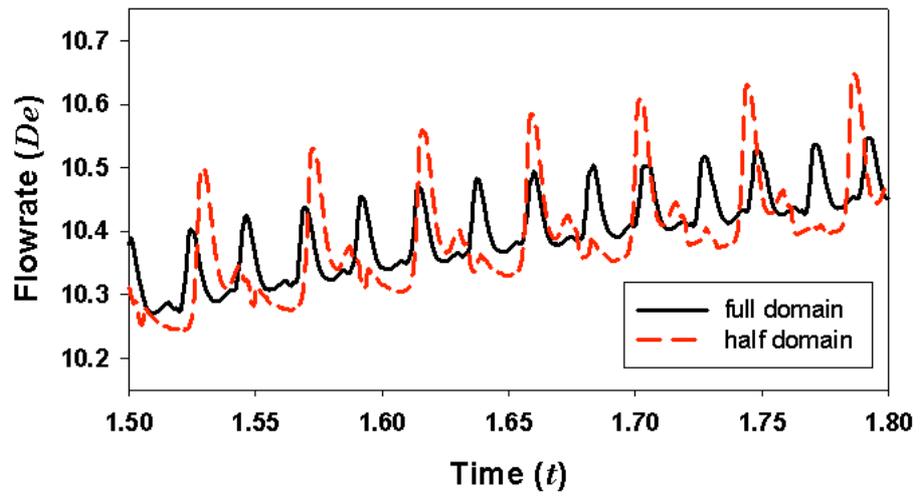

Fig. 19(b)

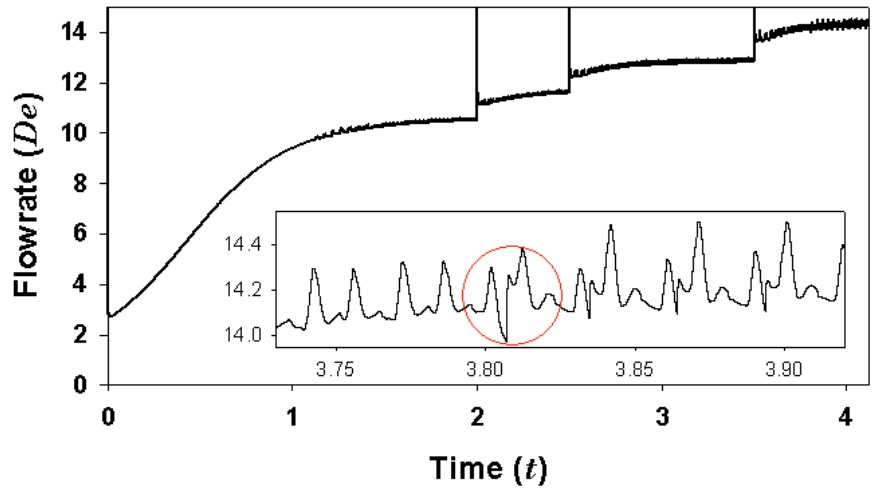

Fig. 20

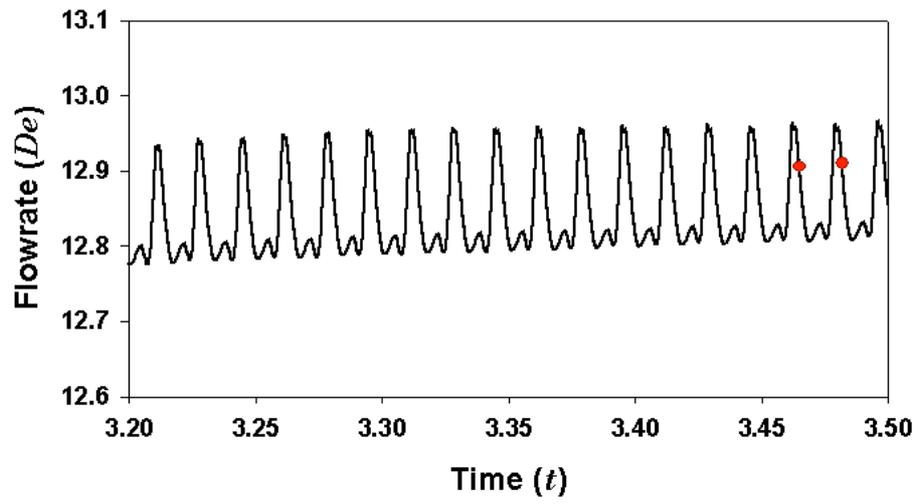

Fig. 21(a)

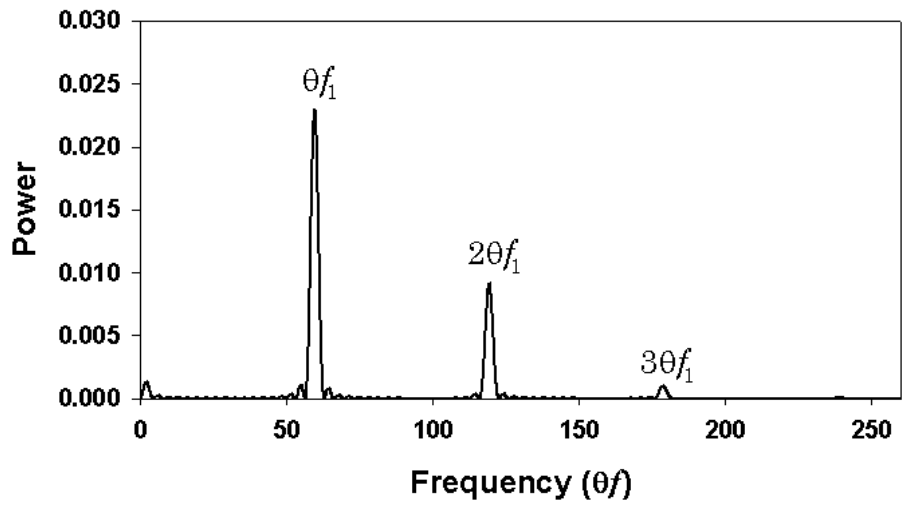

Fig. 21(a)

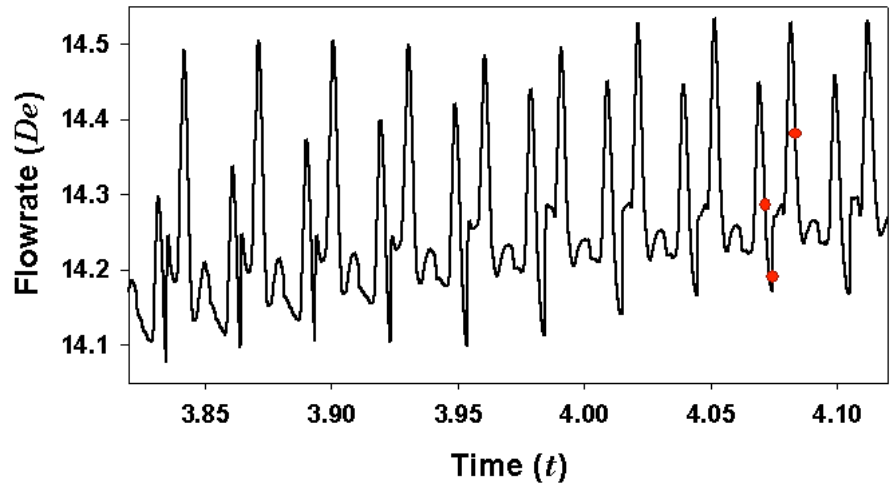

Fig. 21(b)

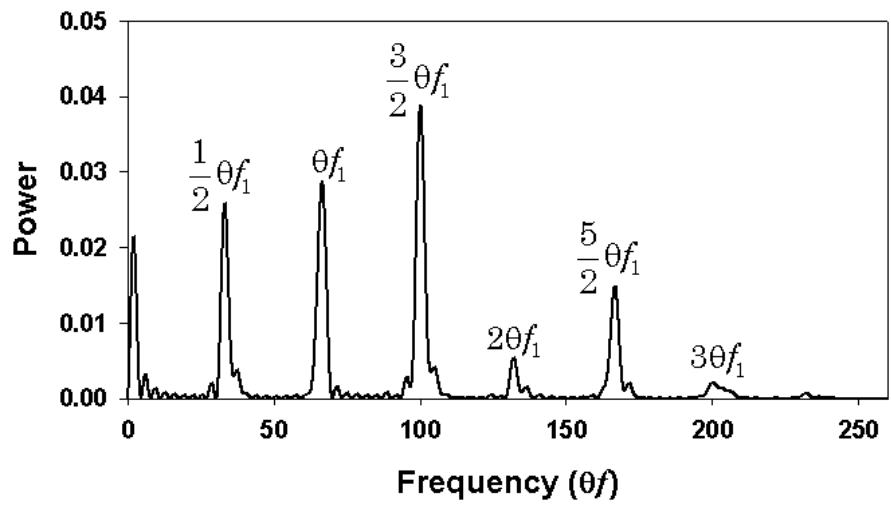

Fig. 21(b)

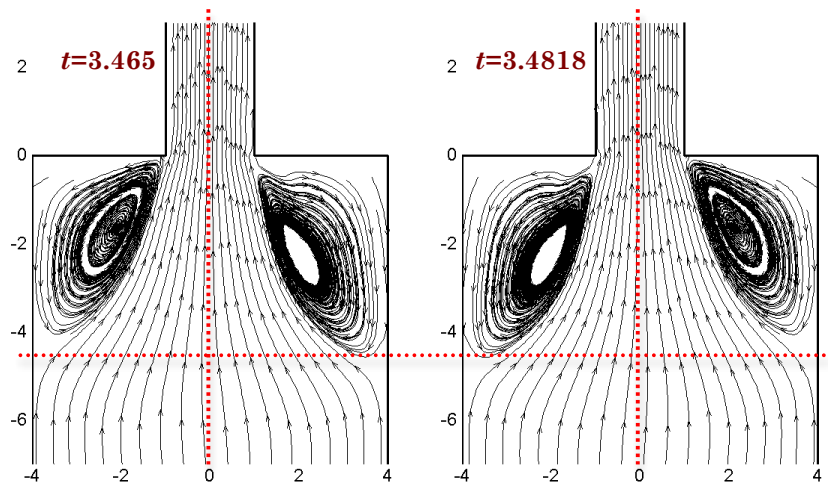

Fig. 22

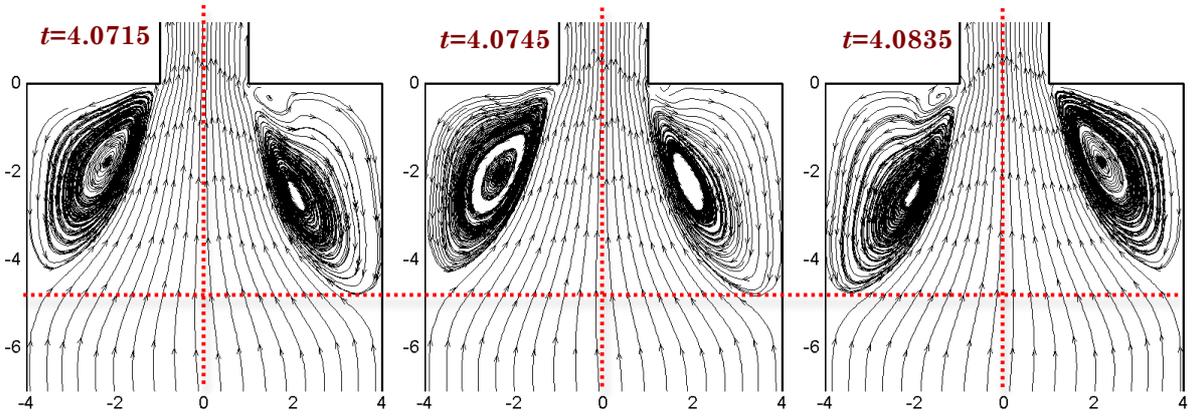

Fig. 23